\documentclass[12pt, amsfonts, amssymb]{article}

\usepackage{amssymb}
\usepackage{amsmath}

\catcode`\@=11
\def\@citex[#1]#2{%
\if@filesw \immediate \write \@auxout {\string \citation {#2}}\fi
\@tempcntb\m@ne \let\@h@ld\relax \def\@citea{}%
\@cite{%
  \@for \@citeb:=#2\do {%
    \@ifundefined {b@\@citeb}%
      {\@h@ld\@citea\@tempcntb\m@ne{\bf ?}%
      \@warning {Citation `\@citeb ' on page \thepage \space undefined}}%
      {\@tempcnta\@tempcntb \advance\@tempcnta\@ne%
      \@tempcntb\number\csname b@\@citeb \endcsname \relax%
      \ifnum\@tempcnta=\@tempcntb 
        \ifx\@h@ld\relax%
          \edef \@h@ld{\@citea\csname b@\@citeb\endcsname}%
        \else%
          \edef\@h@ld{\ifmmode{-}\else--\fi\csname b@\@citeb\endcsname}%
        \fi%
      \else
        \@h@ld\@citea\csname b@\@citeb \endcsname%
        \let\@h@ld\relax%
      \fi}%
    \def\@citea{,\penalty\@highpenalty\,}%
  }\@h@ld
}{#1}}

\def\@citeb#1#2{{[#1]\if@tempswa , #2\fi}}
\def\@citeu#1#2{{$^{#1}$\if@tempswa , #2\fi }}
\def\@citep#1#2{{#1\if@tempswa , #2\fi}}

\def\bcites{         
        \catcode`\@=11
        \let\@cite=\@citeb
        \catcode`\@=12
}

\def\upcites{         
        \catcode`\@=11
        \let\@cite=\@citeu
        \catcode`\@=12
}

\def\plaincites{      
        \catcode`\@=11
        \let\@cite=\@citep
        \catcode`\@=12
}

\catcode`\@=11

\@addtoreset{equation}{section}
\@addtoreset{footnote}{section}
\@addtoreset{footnote}{subsection}

\newcommand{\CC}{{\mathbb C}}
\newcommand{\RR}{{\mathbb R}}
\newcommand{\ZZ}{{\mathbb Z}}
\newcommand{\PP}{{\mathbb P}}

\newcommand{\R}{{\mathbb R}}

\newcommand{\Z}{{\mathbb Z}}
\newcommand{\CP}{{\mathbb C}{\mathbb P}}
\renewcommand{\SS}{{\mathbb S}}
\newcommand{\bP}{{\bar{P}}}
\newcommand{\bD}{{\bar{D}}}
\newcommand{\bz}{{\bar{z}}}

\newcommand{\ra}{\rightarrow}

\newcommand{\nn}{\nonumber}
\newcommand{\LSM}{{\rm LSM}}
\newcommand{\wtX}{\widetilde{X}}

\newcommand{\e}{{\rm e}}
\newcommand{\dd}{{\rm d}}

\title{\bf Worldsheet Descriptions of\\[0.3cm]
Wrapped NS Five-Branes}
\author{
Kentaro Hori \thanks{e-mail: hori@ias.edu}\\
{\it Institute for Advanced Study, Princeton, NJ 08540}
 \rm
\and 
Anton Kapustin\thanks{e-mail: kapustin@theory.caltech.edu}\\
{\it California Institute of Technology, Pasadena, CA 91125}
}

\begin{document}
\begin{titlepage}

\renewcommand{\thepage}{ }

\maketitle

\begin{abstract}
We provide a world-sheet description of Neveu-Schwarz five-branes wrapped on
a complex projective space.
It is an orbifold of the product of
an ${\mathcal N}=2$ minimal model and the IR fixed point of
a certain linear sigma model.
We show how the naked singularity in the supergravity description
is resolved by the world-sheet CFT.
Applying mirror symmetry, we show that
the low-energy theory of
NS5-branes wrapped on $\CC\PP^1$ in Eguchi-Hanson space
is described by
the Seiberg-Witten prepotential for ${\mathcal N}=2$ super-Yang-Mills,
with the gauge group given by the ADE-type of the five-brane.
The world-sheet CFT is generically regular, but singularities develop
precisely at the Argyres-Douglas points and massless monopole points
of the space-time theory.
We also study the low-energy theory of 
NS5-branes wrapped on $\CC\PP^2$ in a Calabi-Yau 3-fold and its relation
to $(2,2)$ super-Yang-Mills theory in two dimensions.

\end{abstract}
\vspace{-7.5in}

\parbox{\linewidth}
{\small\hfill \shortstack{CALT-68-2376}} 

\end{titlepage}

\section{Introduction and Summary}

Finding string theory duals of gauge theories
or gauge theory duals of strings has been regarded as an important 
problem since the work of 't Hooft \cite{tHooft}.
AdS/CFT duality is an important development in this direction.
However, in this case the world-sheet CFT is usually not under control
and one has to work in the supergravity approximation.
Furthermore, one is usually restricted to conformally invariant and
supersymmetric theories.
In particular, we still do not know the string dual of pure Yang-Mills
in $3+1$ dimensions.
It was suggested in Ref.~\cite{Polyakov} that the string dual of
such a theory should be some non-critical string
where the Liouville field represents the renormalization scale.

There is one interesting example of a non-scale-invariant theory
which has a string theory dual
with a controllable worldsheet description: the decoupled theory on
Neveu-Schwarz five-branes \cite{LST}.
When NS5-branes are coincident, the supergravity solution develops an infinite
``throat'' with a linearly growing dilaton \cite{CHS}.
It has been proposed that 
the theory of strings propagating in the throat is dual to the Little
String Theory living on the five-branes~\cite{ABKS}.
The corresponding world-sheet CFT is exactly soluble
and can be used to find the spectrum of the theory.
However, it is not suitable for computing correlators
because the dilaton grows without bound as one moves toward the branes.
When the fivebranes are separated,
the system is still described by the exactly solvable CFT~\cite{OV}:
\begin{equation}
{\mathcal C}_m^{(1)}
=\left[{SL(2,\R)_{m+2}\over U(1)}\times {SU(2)_{m-2}\over U(1)}\right]
/\Z_m,
\label{one}
\end{equation}
where $m$ is the number of five-branes.
$SL(2,\R)_{m+2}/U(1)$ is the Euclidean 2d black hole background
which has a semi-infinite cigar geometry with an asymptotically
linear dilaton~\cite{BH},
while $SU(2)_{m-2}/U(1)$ stands for
the ${\mathcal N}=2$ minimal model.
In this case the dilaton is bounded (its maximal value is achieved at
the tip of the cigar), and one can hope that string perturbation theory
is sufficient. This approach gives interesting results about the
physics of LSTs at finite $m$~\cite{GKP,GK,ES}.

Upon compactification LSTs can flow to a number of supersymmetric
gauge theories in lower dimensions. For example, upon compactification
on $\CC\PP^1$ Type IIB LST flows either to an ${\mathcal N}=2$ or 
${\mathcal N}=1$ super-Yang-Mills
theory in $d=4$, depending on the ``twist.'' 
The twist describes the geometry of the
normal bundle; for example, when the $\CC\PP^1$ is the vanishing cycle
of the Eguchi-Hanson space, one obtains ${\mathcal N}=2$ supersymmetry
 in $3+1$ dimensions.
Clearly, it is of interest to
understand the corresponding superstring backgrounds. The supergravity
solutions describing the near horizon geometries of such wrapped
five-branes have been studied in~\cite{MN,Gauntetal,Bigetal,GR,Gauntetal2}. 
The solution corresponding
to ${\mathcal N}=1$ super-Yang-Mills turns out non-singular
and exhibits a spontaneous breaking of R-symmetry~\cite{MN}.
On the other hand, the supergravity solution corresponding to ${\mathcal N}=2$
super-Yang-Mills has a naked singularity~\cite{Gauntetal,Bigetal}.

In this paper, we provide a regular world-sheet description of
the ${\mathcal N}=2$ compactification.
The world-sheet theory is very similar to that for flat separated
five-branes, except that the two-dimensional black hole in (\ref{one}) is
replaced by a four-dimensional dilatonic background parametrized by $m$.
That is, the world-sheet CFT for $m$ five-branes wrapped on $\CC\PP^1$
is of the form
\begin{equation}
{\mathcal C}_m^{(2)}
=\left[\mbox{4d background}_m\times 
{SU(2)_{m-2}\over U(1)}\right]
/\Z_m.
\label{two}
\end{equation}
The 4d background is the one found in Ref.~\cite{KKL}
by solving the one-loop beta-function equations.
It has a $U(2)$-invariant K\"ahler metric
and a dilaton which is asymptotically linear but is bounded from above.
The system ${\mathcal C}_m^{(2)}$ is a well-defined CFT, whose only
peculiarities come from the non-compactness of the target space. 
We will see that a simple T-duality converts it
to the supergravity solution of Refs.~\cite{Gauntetal,Bigetal}
describing five-branes wrapped on $\CC\PP^1$.
The naked singularity of the latter is thus resolved by
the world-sheet CFT.
In fact, one can describe the system without resort to the one-loop
approximation:
the 4d background can be defined as
the infra-red limit of a gauged linear sigma model (LSM) which is a simple
generalization of the LSM that yields
the 2d black hole~\cite{HK}.

The supergravity solution of Refs.~\cite{Gauntetal,Bigetal} actually has
one parameter $k$ which can take any value from $-1$ to $+\infty.$ 
It was argued to correspond to a Coulomb branch modulus, with
the origin of the Coulomb branch corresponding to $k=-1.$ Likewise, we
are able to construct a one-parameter family of LSMs which corresponds to this
one-parameter family of supergravity solutions. Thus for any $k$ the
naked singularity is resolved by world-sheet CFT. There is one exception
though: at a single point on this line the world-sheet CFT becomes singular.
Using the mirror description (see below), we show
that at the singularity the theory on the five-branes flows to an
${\mathcal N}=2$ $d=4$ SCFT of the Argyres-Douglas type.

Unlike the Euclidean 2d black hole, its 4d analogue does not appear
to be integrable. Nevertheless, when the number of five-branes $m$ is large,
it is weakly curved everywhere, and thus all scattering amplitudes
can, in principle, be computed order by order in $1/m$ expansion. 
Combined with the 
exactly known correlators of the minimal model, they provide a complete
holographic description of five-branes wrapped on $\CC\PP^1$.

Linear sigma models that appear in this paper
have mirror duals of the Liouville type
\cite{HV,HK}.
Combined with the fact that the orbifold CFT ${SU(2)_{m-2}\over U(1)}/\Z_m$
is mirror to the LG model with the superpotential $W=X^m$, this allows us
to find the mirror descriptions of the above CFTs.
The mirror of the CFT (\ref{one})
is the Landau-Ginzburg orbifold
\begin{equation}
{\mathcal C}_m^{(1)}\stackrel{\rm mirror}{=}
\left[W=\e^{-mZ}+X^m\right]/\Z_m,
\label{mirone}
\end{equation}
where $\Z_m$ action is generated by
$Z\to Z-{2\pi i\over m}$ and $X\to \e^{2\pi i\over m}X$.
Moving NS5-branes corresponds to deforming the superpotential
by adding the terms $\e^{-(m-\ell)Z}X^{\ell}$, $\ell=0,1,...,m-2$.
The relation (\ref{mirone}) entails the relation between NS5-branes and ALE spaces proposed in \cite{OV} .
The family of deformations $W=\mu\e^{-mZ}+X^m$ contains a singular
point $\mu=0$ which corresponds to coincident NS5-branes, or
equivalently the singular ALE space.

The mirror of the CFT (\ref{two}) for NS5-branes wrapped on $\CC\PP^1$
is
\begin{equation}
{\mathcal C}_m^{(2)}\stackrel{\rm mirror}{=}
\left[W=\e^{-mZ}(\e^{-Y}+\e^Y)+X^m\right]/\Z_m,
\end{equation}
where $\Z_m$ acts on $Z$ and $X$ in the same way as above and leaves
$Y$ invariant. Moving on the Coulomb branch of the five-brane theory
again corresponds to adding
the terms $\e^{-(m-\ell)Z}X^{\ell}$, $\ell=0,1,...,m-2$.
Using this mirror description, one can reproduce
the Seiberg-Witten solution of
the low energy ${\mathcal N}=2$ super-Yang-Mills theory.
The family of deformations $W=\e^{-mZ}(\e^{-Y}+\e^Y+u_0)+X^m$
corresponds to the one parameter family of
supergravity solutions mentioned above.
We will see that the world-sheet theory is singular at $u_0=\pm 2$.
{}From the point of view of low-energy super-Yang-Mills theory, this 
is the Argyres-Douglas point. As in the case of the conifold singularity, 
the breakdown of the world-sheet CFT signals non-trivial infrared physics 
in target space.

In this paper, we will also consider a
compactification of LST on $\CC\PP^2$ which preserves $(2,2)$ supersymmetry 
in $1+1$ dimensions.
This corresponds to NS5-branes wrapped on a $\CC\PP^2$ in a non-compact 
Calabi-Yau 3-fold, the total space of the line bundle ${\mathcal O}(-3)$ over $\CC\PP^2$. The world-sheet CFT is of the form
\begin{equation}
{\mathcal C}_m^{(3)}=\left[\mbox{6d background}_m\times
{SU(2)_{m-2}\over U(1)}\right]/\Z_m,
\end{equation}
where the 6d background can be described as
a $U(3)$-invariant dilatonic solution
of the one-loop beta function equations for large $m$,
or as the infra-red limit of a simple LSM for any $m$.
The LSM realization enables us to find the mirror theory,
which is closely related to the world-sheet CFT of
a certain non-compact Calabi-Yau 4-fold. The mirror description
enables us to compute world-sheet instanton correction
to the space-time superpotential.

In the infrared, LSTs compactified on complex projective spaces are in the 
same universality class as super-Yang-Mills theories in the remaining 
flat dimensions, but are not quite the same as the latter.
One could hope to find some limit where extra degrees of freedom decouple
and one obtains exactly super-Yang-Mills.
Unfortunately, we find that in the decoupling limit the string coupling 
should be taken large, 
both for $\CC\PP^1$ and $\CC\PP^2$ compactifications.
Thus our worldsheet descriptions are applicable only
when the compactified LST is quite different from the
super-Yang-Mills theory.
However, in the $\CC\PP^1$ case,
due to a supersymmetric non-renormalization theorem,
the low energy prepotential
does not depend on the string coupling, and we recover the correct solution
of the super-Yang-Mills theory. 
In fact, world-sheet intantons precisely correspond to Yang-Mills
instantons, which are
responsible for non-perturbative corrections to the prepotential.
On the other hand, for the $\CC\PP^2$ compactification
we argue that and world-sheet instanton contributions
to the space-time (twisted) superpotential vanish in the 
decoupling limit.

The paper is organized as follows. In Section~\ref{sec:twist} we review
``twisted'' compactifications of five-branes on $\CC\PP^n$ and their
symmetries. In Section~\ref{sec:LSM} we define linear sigma models of
interest to us and study their infrared fixed points in world-sheet perturbation
theory. In Section~\ref{sec:T} we relate these linear sigma models to
wrapped five-branes. In Section~\ref{sec:mirror} we find the mirror
description of our world-sheet CFTs in terms of Landau-Ginzburg models.
In Section~\ref{sec:CP1} we use this mirror description to determine
the Seiberg-Witten prepotential for five-branes wrapped on $\CC\PP^1$
and investigate the world-sheet interpretation of Argyres-Douglas 
points. We also obtain the supergravity solution describing five-branes
away from the origin of the Coulomb branch by studying the RG flow on the
world-sheet. In Section~\ref{sec:CP2} we perform a similar analysis for
five-branes wrapped on $\CC\PP^2$. Section~\ref{sec:decoup} discusses the
decoupling limit in which the theory of wrapped five-branes reduces
to super-Yang-Mills. We conclude with remarks about the (lack of) 
integrability of the world-sheet CFT, connection with non-critical
superstrings, and other issues.

Throughout this paper we set $\alpha'=1$. In particular, the gauge
coupling of the low energy Yang-Mills theory on NS5-branes is given by
$g_{6}^2=2(2\pi)^3$.

\section{Five-Branes Wrapped on $\CC\PP^n$}\label{sec:twist}

Type IIB NS5-brane configurations in flat space
which preserve sixteen supercharges have an ADE classification.
Simply putting $m$ five-branes on top of each other yields
a configuration of type $A_{m-1}.$ If $m$ is even,
then putting $\frac{m}{2}$ five-branes on top of an $ON^-$ six-plane yields
a configuration of type $D_{\frac{m}{2}+1}.$ We remind that an $ON^-$ plane
is defined as an orbifold of Type IIB string by the symmetry
which reflects four of the coordinates and in addition multiplies all
fields by $(-1)^{F_L},$ where $F_L$ is the left-moving fermion number.
The near-horizon limit is given by the Callan-Harvey-Strominger
background \cite{CHS}
\begin{equation}
\R^{5+1}\times\R\times \SS^3,
\end{equation}
where the dilaton depends linearly on the affine coordinate on $\R$
and there is an H-flux through $\SS^3$. Thus the $\SS^3$ part is described
by an $SU(2)$ WZW model.
The ADE classification of five-branes originates from the ADE classification of
affine $SU(2)$ modular invariants. 
Each such brane configuration gives rise to
a Little String Theory with sixteen supercharges in $d=6$.
The $SU(2)_{\ell}\times SU(2)_r$ current algebra of the $SU(2)$ WZW model
gives rise to the R-symmetry group of the LST. 
Under the $SO(1,5)\times SU(2)_{\ell}\times SU(2)_r $
bosonic symmetry the supercharges 
transform as $({\bf 4_+},{\bf 2},{\bf 1})+
({\bf 4_-},{\bf 1},{\bf 2}),$ and satisfy a reality condition.

When compactifying on $\CC\PP^1,$ we can ``twist'' the theory by embedding
the $U(1)$ structure group of the $\CP^1$ spin bundle 
into the R-symmetry group. We choose to embed $U(1)$ {\it diagonally} into
the $U(1)_{\ell}\times U(1)_r$ subgroup of the
$SU(2)_{\ell}\times SU(2)_r$ R-symmetry.
In this case each of the two supercharge multiplets in the $d=6$ theory yields
a complex supercharge in $d=4$. Therefore we end up with an ${\mathcal N}=2$
theory in $d=4$.
The $U(1)$ part of the 
$SU(2)\times U(1)$
R-symmetry group is the $U(1)$
which is {\it off-diagonally} embedded into
$U(1)_{\ell}\times U(1)_r$.
The $SU(2)$ R-symmetry of the low-energy theory is not manifest,
except the $U(1)$ subgroup which can be identified as the diagonal
of $U(1)_{\ell}\times U(1)_r$.
We note that the $d=6$
R-symmetry group $SU(2)_{\ell}\times SU(2)_r$
acts on the group element of the $SU(2)$ WZW model as
$g\mapsto g_{\ell}gg_r^{-1}$.
Thus the $U(1)$ R-symmetry group in $d=4$ acts as follows:\footnote{
We realize $SU(2)$ as the group of unit quaternions.}
\begin{equation}
U(1)_R:g\mapsto \e^{i\alpha}g\e^{i\alpha},
\label{RU1}
\end{equation}
while the action of the $U(1)$ subgroup used in the twisting is
\begin{equation}
U(1)_{\it twist}:
g\mapsto \e^{i\gamma}g\e^{-i\gamma}.
\label{twistU1}
\end{equation}
This twist is the same as in Ref.~\cite{Gauntetal,Bigetal}.
\footnote{The twist used in \cite{MN} for ${\mathcal N}=1$ compactification
on $\CP^1$
is $g\mapsto \e^{i\gamma}g$.}
As explained in Ref.~\cite{Gauntetal}, it
corresponds to five-branes wrapped on the holomorphic 2-cycle of the
Eguchi-Hanson space.

In the low-energy limit an LST of type $A,D,$ or $E$ reduces to
${\mathcal N}=2$ $d=6$ super-Yang-Mills theory with a simply laced gauge group
of the appropriate type. It is natural to expect that its ``twisted''
Kaluza-Klein reduction on $\CC\PP^1$ yields ${\mathcal N}=2$ $d=4$
super-Yang-Mills. Indeed, it is clear that we get an ${\mathcal N}=2$
vector multiplet in $d=4.$ There is a single complex scalar in such
a multiplet. On the other hand, the scalars in $d=6$ transform as a $(2,2)$
of the R-symmetry group, and therefore upon Kaluza-Klein reduction they
yield precisely one complex scalar. This leaves no room for any 
hypermultiplets.

Similarly, we can consider a twisted compactification of Type IIB LST
on $\CP^2$.
Although $\CP^2$ does not admit a spin structure,
there is a twisted compactification that is well defined and preserves
four supercharges.
We recall that the tangent bundle
of a K\"ahler surface has a reduced structure group
$U(2)\subset SO(4)$, and its lift to the spin group $SU(2)\times SU(2)$
is $SU(2)\times U(1)$. The latter group covers $U(2)$ twice, with kernel
$(-1,-1)$.
The twisting we choose is to embed the spinor group $SU(2)\times U(1)$
into the $SU(2)_{\ell}\times SU(2)_r$ R-symmetry group,
so that the $SU(2)$ factor is embedded trivially, while
the $U(1)$ factor is embedded diagonally
into $U(1)_{\ell}\times U(1)_r$.
Then the $d=6$ supercharges transform {\it vectorially} under
the twisted $SU(2)\times U(1)$ action
--- the kernel $(-1,-1)$ acts trivially.
In other words, the modified spin group is $U(2)$.
Under this modified spin group, the supercharges transform as follows:
\begin{equation}
\begin{array}{l}
({\bf 4}_+,{\bf 2},{\bf 1})=
{\bf 2}_++\overline{\bf 2}_++{\det}_-+\overline{\det}_-+{\bf 1}_-+{\bf 1}_-,
\\
({\bf 4}_-,{\bf 1},{\bf 2})=
{\bf 2}_-+\overline{\bf 2}_-+{\det}_++\overline{\det}_++{\bf 1}_++{\bf 1}_+.
\end{array}
\label{superU}
\end{equation}
On the right hand side,
${\bf 2}$, $\overline{\bf 2}$, $\det$, $\overline{\det}$, ${\bf 1}$
are the fundamental, the anti-fundamental, determinant, anti-determinant,
and trivial representations of $U(2)$ respectively
(the subscripts $\pm$ show the chirality in
the uncompactified $1+1$ dimensions).
In particular, supercurrents are sections of well-defined vector bundles
on $\CP^2$.

To prove that this twisted compactification is well-defined, we need to 
show that {\it all} fields of the LST transform vectorially
under the twisted spinor group.
Since the supercharges are well-defined as we have seen, it is enough to
show that bosons transform vectorially.
This is equivalent to showing that bosons transfom
vectorially under the $SU(2)_{\ell}\times SU(2)_r$ R-symmetry. In other
words, we need to check that $j_{\ell}+j_r\in \Z$.
The spectrum of LST
can be found using the holographic description \cite{ABKS}.
In particular, the $SU(2)_{\ell}\times SU(2)_r$ spins for bosons
are the same as in the $SU(2)$ WZW model up to integer shifts.
For all ADE modular invariants, the condition $j_L+j_R\in\Z$ is satisfied
\cite{CIZ}. This completes the proof.

Since the holonomy group of $\CP^2$ is the entire
$U(2)$, Eq.~(\ref{superU})  shows that
there are only four covariantly
constant spinors on $\CP^2$ --- two ${\bf 1}_+$'s and two ${\bf 1}_-$'s.
Thus we end up with $(2,2)$ supersymmetric field theory in
$1+1$ dimensions.
The $U(1)_{\ell}$ and $U(1)_r$ R-symmetries descend to the
left and right R-symmetries of the $(2,2)$ algebra.
It is also easy to see that the twisted Kaluza-Klein reduction
of an ${\mathcal N}=(1,1)$ $d=6$ vector multiplet yields a single
$(2,2)$ vector multiplet in $d=2$.
Thus the na\"\i ve dimensional reduction of $(1,1)$ LST
yields $(2,2)$ super-Yang-Mills theory in $1+1$ dimensions,
with an ADE gauge group.

Note that this
twist corresponds to $\CC\PP^2$ being an exceptional divisor
in a Calabi-Yau 3-fold.
The local model for this is an $O(-3)$ bundle over $\CC\PP^2.$
There exists an ALE Ricci flat K\"ahler metric on this manifold which
generalizes the Eguchi-Hanson metric~\cite{Calabi}, but we will not
need its explicit form here. Another way to think about this non-compact
Calabi-Yau is as a crepant resolution of the $\CC^3/\ZZ_3$ orbifold
singularity~\cite{Calabi}.

To decouple Kaluza-Klein modes one must take the size of the compactification
manifold to be small. Since LSTs are non-local theories, it is not
completely clear what this means. In other words, it is not obvious in
what regime one can simultaneously decouple the massive Kaluza-Klein modes
and approximate the LST by $d=6$ super-Yang-Mills. This issue is discussed
in more detail in Section~\ref{sec:decoup}.

\section{Linear Sigma Model and Its Infrared Limit}\label{sec:LSM}

In this section we study a certain gauged linear sigma model. In the next section, we will tensor it with an ${\mathcal N}=2$
minimal model and show that it is T-dual to a configuration of wrapped
five-branes.

We follow the notations and conventions of Ref.~\cite{HK}. 
We consider an ${\mathcal N}=2$ LSM with the following superspace Lagrangian:
\begin{equation}
S=\frac{1}{2\pi} \int d^2x d^4\theta\left[\sum_{i=1}^n \Phi_i^\dag e^V \Phi_i+
\frac{k}{4}\left(P+\bP + V\right)^2 -\frac{1}{2e^2} \Sigma^2\right].
\label{LSMnk}
\end{equation}
Here $P$ and $\Phi_i,i=1,\ldots,n,$ are chiral superfields, $V$ is an
abelian vector superfield, and $\Sigma=\bD_+ D_- V$. Under $U(1)$ gauge
transformations the fields transform as
$\Phi_i\to\e^{i\Lambda}\Phi_i$, $P\to P+i\Lambda$ and
$V\to V-i\Lambda+i\overline{\Lambda}$,
where $\Lambda$ is a chiral superfield.
The imaginary part of $P$ is periodically
identified with period $2\pi.$

The chiral field $P$ can be dualized to a twisted chiral superfield
$Y_P$. In terms of $Y_P,\Phi_i,$ and $V$ the Lagrangian takes the form
$$
S=\frac{1}{2\pi} \int d^2x d^4\theta\left[\sum_{i=1}^n \Phi_i^\dag e^V\Phi_i-
\frac{1}{2k}\vert Y_P\vert^2 -\frac{1}{2e^2} \Sigma^2\right]+
\frac{1}{4\pi}\left(\int d^2x d^2\tilde{\theta} \Sigma Y_P + h.c.\right).
$$

This ${\mathcal N}=2$ gauge theory has a non-anomalous axial 
$U(1)_R$ symmetry,
therefore in the infrared the theory flows to a non-trivial CFT. 
The central charge of this CFT has been computed in Ref.~\cite{HK}:
\begin{equation}\label{LSMcentralc}
c_{IR}=3n\left(1+\frac{2n}{k}\right).
\end{equation}
This result is exact. For large $k$ the CFT is weakly coupled and
can be represented by an
${\mathcal N}=2$ supersymmetric non-linear
sigma model with a $2n$-dimensional
target space. Since there is no $H$-field in this CFT, the target
space must be K\"ahlerian.

Our goal in this section is to determine this K\"ahlerian metric, at least
for large $k$, where one can use sigma-model perturbation theory.
Following Ref.~\cite{HK}, we do it in two steps. On the first step,
we integrate out the gauge fields classically, which is a good approximation
for large $k$. This results in a non-linear sigma model with a $2n$-dimensional
K\"ahlerian target space. This target space, however,
does not solve the beta-function equations. On the second
step, we find the K\"ahlerian $2n$-dimensional metric which solves the beta-function equations and has the same symmetries and asymptotics as 
the metric found in step one. Since such a metric is unique, the LSM must 
flow to the non-linear sigma-model with this metric. 
Note that the one-loop beta function equations
are sufficient because both the initial and final K\"ahlerian metrics
are weakly curved for large $k$. 

\vskip20pt
\noindent {\it Step 1}

\noindent In principle, one could integrate out the vector multiplet 
in superspace.
However, the resulting K\"ahler potential is not expressible in elementary
functions, and therefore we prefer to work with component fields.
For simplicity, we provide a detailed derivation for the case $n=2$
and then write down the answer in the general case.

The kinetic energy for the bosonic matter fields is encoded in the 
following metric:
$$
\dd s^2=2|\dd \Phi_1|^2+2|\dd\Phi_2|^2+k|\dd P|^2.
$$

It is convenient to parametrize $\Phi_i$ as follows:
\begin{eqnarray}
\Phi_1&=&re^{\frac{i}{2}(\psi-\phi)}\cos\frac{\theta}{2},\\
\Phi_2&=&re^{\frac{i}{2}(\psi+\phi)}\sin\frac{\theta}{2},
\end{eqnarray}
where $r\in [0,+\infty),$ $\theta\in [0,\pi],$ $\phi\sim \phi+2\pi,$
$\psi\sim \psi+4\pi.$ This amounts to regarding $\Phi_1,\Phi_2$ as
coordinates on the total space of the tautological line bundle
on $\CC\PP^1$ parametrized by $\theta$ and $\phi.$ The coordinates
$r,\psi$ parametrize the fiber of this line bundle.

In the new coordinates the metric takes the form
$$
\dd s^2=2\dd r^2+\frac{r^2}{2}\left((\dd\psi-\cos\theta \dd\phi)^2+
\dd\theta^2+\sin^2\theta \dd\phi^2\right)+k |\dd P|^2.
$$

The D-term constraint reads
$$
r^2+k\ {\rm Re}\,P=0.
$$
It allows to eliminate (i.e. integrate out) ${\rm Re}\,P$. 

To eliminate ${\rm Im}\,P$, we gauge the $U(1)$ isometry 
$$P\ra P + i\alpha,\quad \psi\ra \psi+2\alpha,$$
and then gauge away ${\rm Im}\,P.$ Finally, we integrate
out the gauge field. 

The resulting four-dimensional metric has the form
$$
\dd s^2=2f(r) \dd r^2+\frac{r^2}{2f(r)}
\left(\dd\psi-\cos\theta \dd\phi\right)^2+
\frac{r^2}{2}\left(\dd\theta^2+\sin^2\theta \dd\phi^2\right),
$$
where 
$$
f(r)=1+\frac{2r^2}{k}.
$$

This metric is everywhere smooth and weakly curved for large $k$.
By construction, it is guaranteed to be K\"ahlerian. 
The fermionic terms in the Lagrangian are uniquely fixed by the
requirement of ${\mathcal N}=2$ supersymmetry.

For general $n$ the result is similar. The target space is topologically
a $\CC^n$, but it is best regarded as the total space of the tautological
line bundle on $\CC\PP^{n-1}$. Let $A$ be the connection 1-form of the
natural connection on this line bundle. (Its curvature is the standard
K\"ahler form on $\CC\PP^{n-1}$). Then the metric has the form
\begin{equation}\label{narbUVmetric}
\dd s^2=2\dd r^2 f(r) +
\frac{2r^2}{n^2f(r)}\left(\dd\psi-nA\right)^2 +2r^2 ds^2_{FS}.
\end{equation}
Here the function $f(r)$ is the same as above, $ds^2_{FS}$ is the Fubini-Study metric on $\CC\PP^{n-1}$, and $A$ is a connection 1-form on $\CC\PP^{n-1}$
whose curvature is the Fubini-Study K\"ahler form. The variable
$\psi$ has period $2\pi n$. On the affine chart
$\CC^{n-1}\subset \CC\PP^{n-1}$ the metric $ds^2_{FS}$ and the 1-form
$A$ are given by
\begin{eqnarray}
ds^2_{FS}&=&\left(1+\sum_{k=1}^{n-1} |z_k|^2\right)^{-1} 
\sum_{i,j=1}^{n-1} \left(\delta_{ij}-\frac{z_i\bz_j}{1+\sum_k |z_k|^2}
\right) d\bz_i dz_j,\\
A&=& \frac{-i}{2}\frac{1}{1+\sum_k |z_k|^2} \sum_{i=1}^{n-1}
\left(z_i d\bz_i - \bz_i dz_i\right).
\end{eqnarray}

Since the original LSM has $U(n)$ flavor symmetry, the metric of
the non-linear sigma model must have a $U(n)$ isometry. This isometry is manifest in the above expression.

\vskip20pt
\noindent{\it Step 2}

\noindent
Now we need to find a solution to the one-loop beta-function equations 
which is K\"ahlerian, has a $U(n)$ isometry, and asymptotes 
to the metric found in Step 1. The technology for doing this has
been developed in Ref.~\cite{KKL}. After correcting some minor inaccuracies
in Ref.~\cite{KKL}, we find that in the case $n=2$
there is a unique metric satisfying all these requirements:
\begin{equation}\label{ntwoIRmetric}
\dd s^2
=\frac{g_2(Y)}{2}\dd Y^2
+\frac{1}{2g_2(Y)}\left(\dd\psi-\cos\theta \dd\phi\right)^2
+\frac{Y}{2}\left(\dd\theta^2+\sin^2\theta \dd\phi^2\right),
\end{equation}
where 
$$
g_2(Y)=\frac{8}{k^2}\frac{Y}{e^{-4Y/k}-1+\frac{4Y}{k}}.
$$
The corresponding dilaton is given by
\begin{equation}\label{ntwoIRdilaton}
\Phi=-\frac{2Y}{k}+const.
\end{equation}
The variable $Y$ runs from $0$ to $+\infty,$ the variable $\psi$ has period $4\pi,$
while $\theta$ and $\phi$ are the usual coordinates on a unit 2-sphere.

The central charge can be found from
$$
c=6+3\left(2(\nabla\Phi)^2-\nabla^2\Phi\right),
$$
and comes out to be
$$
6\left(1+\frac{4}{k}\right).
$$
This agrees with Eq.~(\ref{LSMcentralc}) and therefore provides a check on 
our computation.

The metric Eq.~(\ref{ntwoIRmetric}) is everywhere smooth and weakly
curved for large $k$. It agrees with Eq.~(\ref{ntwoIRmetric}) in
the region of large $Y$ (the corresponding change of variables is
$Y=r^2$).

For general $n$ the result is similar. The metric and the dilaton are
given by
\begin{eqnarray}\label{genn}
\dd s^2&=&\frac{g_n(Y)}{2} \dd Y^2 + 
\frac{2}{n^2 g_n(Y)}\left(\dd\psi-nA\right)^2 +2Y \dd s^2_{FS}\\
\Phi&=&-\frac{nY}{k},
\end{eqnarray}
where the function $g_n(Y)$ is given by
$$
g_n(Y)=\frac{Y^{n-1}}{n} \frac{e^{2nY/k}}{\int_0^Y t^{n-1} e^{2nt/k} dt}.
$$ 
The range of the variable $Y$ is the same as before, while $\psi$
has period $2\pi n$.
This is a unique K\"ahlerian metric which is everywhere smooth, has a 
$U(n)$ isometry, solves the beta-function equations of motion, and 
asymptotes to the metric (\ref{narbUVmetric}) for $Y\ra\infty$. 
The central charge of this supergravity solution can be computed from
$$
c=3n+3\left(2(\nabla\Phi)^2-\nabla^2\Phi\right).
$$
It can be easily checked that it agrees with Eq.~(\ref{LSMcentralc}).

\section{T-Dual Background and Wrapped Branes}\label{sec:T}

First, let us orbifold the background Eq.~(\ref{genn}) by the $\ZZ_n$ symmetry
which acts on the target space coordinates by $\psi\ra\psi + 2\pi.$ 
The action on the fermions is fixed by the requirement
that this symmetry commute with world-sheet supersymmetry.
On the level of the LSM, we gauge the following discrete symmetry:
$$
\Phi_i\ra \exp\left(\frac{2\pi i}{n}\right)\Phi_i, \quad i=1,\ldots,n.
$$

Second, we set $n=2$ and tensor the above dilatonic background with an 
${\mathcal N}=2$ minimal model \cite{SKY}.
In the supergravity
approximation the latter can be thought of as the sigma model whose
target space is the disc
$\{z=\e^{i\beta}\cos\rho\}$
with the metric
$$
\dd s^2=m\left(d\rho^2+\cot^2\rho\ d\beta^2\right), \quad
\quad m=2,3,\ldots,
$$
and the dilaton
$$
\Phi=-\log\sin\rho+const.
$$
There is a genuine curvature singularity at the boundary of the disc,
$\rho=0$, where the dilaton blows up. However, this is
simply due to a bad choice of description and
a T-duality makes it regular there \cite{MMS}.
This supergravity solution has an obvious $U(1)$ isometry.
But the corresponding CFT preserves only its $\ZZ_m$ subgroup.
The breaking can be attributed to world-sheet instanton effects.

Finally, we take a $\Z_m$ orbifold of the product CFT to eliminate 
fractional R-charges.
The relevant orbifold group is generated by the combination of
$\psi\to\psi+2\pi/m$ and the generator of
the $\Z_m$ symmetry of the minimal model.
Since we have taken the $\Z_2$ orbifold at the first step,
the total orbifold group is $\Z_{2m}$.
As usual, the orbifold theory has a quantum $\Z_{2m}$ symmetry,
which we denote $\Z_{2m}^{\rm qua}$.

The central charge of the minimal model is
$$
c_{min}=3\left(1-\frac{2}{m}\right), \quad m=2,3,\ldots.
$$
Thus if we set $k=4m,$ the total central charge is
$$
c_{tot}=3\left(1-\frac{2}{m}\right)+6\left(1+\frac{1}{m}\right)=9.
$$
Hence we can obtain a bona fide superstring background with $c=15$
by throwing in two more free chiral superfields.

The final result is an
${\mathcal N}=2$ CFT which describes a Type II superstring background with
four-dimensional Poincar\'{e} invariance and an additional $U(2)$ symmetry.
${\mathcal N}=2$ superconformal symmetry on the world-sheet gives rise to
${\mathcal N}=2$ $d=4$ supersymmetry in target space in the usual manner.
The quantum symmetry $\Z_{2m}^{\rm qua}$ on the worlsheet
corresponds to a $\ZZ_{2m}$ R-symmetry in target space, as we will see
momentarily.

To establish a connection with NS five-branes wrapped on $\CC\PP^1$,
we perform a T-duality along a Killing vector field
\begin{equation}
v_K=\frac{1}{m}\left(\frac{\partial}{\partial \beta}
+\frac{\partial}{\partial \psi}\right).
\label{Killing}
\end{equation}
Note that this vector field generates a circle action of period $2\pi$
because of the $\Z_m$ orbifold on the last step.
Thus, the dual variable, which we denote by $\lambda,$ will
also have period $2\pi$.
The T-dual background has the following form:
\begin{eqnarray}\label{SUGRAP1}
\dd s^2&=& \frac{g_2(Y)}{2} \dd Y^2+
\frac{Y}{2}\left(\dd\theta^2+\sin^2\!\theta \dd\phi^2\right)
+m\dd\rho^2\\
&+& \left(1+2m\cot^2\!\rho\ g_2(Y)\right)^{-1}
\left( m\cot^2\!\rho\ (\dd\psi-\cos\theta \dd\phi)^2 + 
2m^2 g_2(Y) \dd\lambda^2\right),\nn\\
B&=& \frac{m}{2\pi} \frac{\dd\lambda\wedge  
\left(\dd\psi-\cos\theta \dd\phi\right)}{1+2m\cot^2\!\rho\ 
g_2(Y)} \\
\Phi&=&-\frac{Y}{2m}-\frac{1}{2}\log\left(\frac{m}{2}\cos^2\!\rho+
\frac{\sin^2\!\rho}{4g_2(Y)}\right)+const.
\end{eqnarray}
The function $g_2(Y)$ is given by
$$
g_2(Y)=\frac{1}{2m}\frac{Y/m}{e^{-Y/m}-1+\frac{Y}{m}}.
$$

The T-dual background has a curvature singularity at the locus
\begin{equation}
Y=0,\quad \rho={\pi\over 2}.
\end{equation}
The dilaton and the norm of H-field also diverge there.
At large $Y$, the function $g_2(Y)$ approaches $1/2m,$ and
the solution asymptotes to $\Phi\sim -Y/2m$ and
\begin{eqnarray}
\dd s^2&\sim&\frac{1}{4m}\dd Y^2+
{Y\over 2}\left(\dd\theta^2+\sin^2\!\theta \dd\phi^2\right)
\nn\\
&&+m\left[\dd\rho^2+\cos^2\!\rho (\dd\psi-\cos\theta\dd\phi)^2
+\sin^2\!\rho\,\dd\lambda^2\right],\\
B&\sim&{m\over 2\pi}\sin^2\!\rho\,\dd\lambda\wedge (\dd\psi-\cos\theta\dd\phi).
\end{eqnarray}
For a fixed value of $(\theta,\phi)$, the metric for
$(\rho,\psi,\lambda)$ is that of a round $\SS^3$ realized by
$|z_1|^2+|z_2|^2=m$, where
$z_1=\sqrt{m}\e^{i\lambda}\sin\rho$ and
$z_2=\sqrt{m}\e^{i\psi}\cos\rho$.
We also note that the $H$-flux through any such $\SS^3$ is
$2\pi m$.
Thus we have an $SU(2)$ WZW model fibered over $\CC\PP^1$,
where the fibration is determined by the combination
$(\dd\psi-\cos\theta\dd\phi)$ that appears
in the solution.
Note that $A=\cos\theta\dd\phi$ is the Levi-Civita
connection of $\CP^1$
(and ${1\over 2}A$ is the connection on the spin bundle $Spin(\CP^1)$).
Note also that the
shift $\psi\to \psi-2\gamma$ corresponds to
the $U(1)_{\it twist}$ action (\ref{twistU1}) for
the $\SS^3$ coordinates $g=z_1+jz_2$.
that is used in the ${\mathcal N}=2$ twisting.
Thus at large $Y$ the geometry is that of an $SU(2)$ bundle
over $\CP^1$ associated with a circle bundle with unit Euler number
via the action of $U(1)_{\it twist}.$ 
This shows that we are dealing with the ${\mathcal N}=2$ compactification of
$m$ NS5-branes on $\CC\PP^1.$

The $U(1)_R$ action (\ref{RU1}) is
$z_1+jz_2\to \e^{i\alpha}(z_1+jz_2)\e^{i\alpha}$
and hence is a shift $\lambda\to\lambda+2\alpha$.
It measures twice the momentum in $\lambda$-direction.
In the original description (before T-duality), it measures
twice the winding in the direction $U(1)_K$ given by the Killing vector
field $v_K$ in (\ref{Killing}).
However, a string
which is wound $2m$ times around $U(1)_K$
lifts to a well-defined configuration in the geometry
before the $\Z_{2m}$ orbifold.
Such a configuration is topologically trivial since
the total geometry before orbifolding has no non-contracable loop.
Thus, the winding number in $U(1)_K$ is only conserved modulo $2m$.
(The conserved charge is that of the quantum symmetry $\Z_{2m}^{\rm qua}$.)
In other words, momentum in $\lambda$-direction is conserved only modulo
$2m$.
This corresponds to the anomaly of the $U(1)_R$ symmetry
\begin{equation}
U(1)_R\to\Z_{4m}.
\end{equation}
We have shown, as promised, that (the $\Z_2$-quotient of) this R-symmetry
is nothing but the quantum symmetry
$\Z_{2m}^{\rm qua}$.
On the other hand, the $U(1)$ subgroup of the $SU(2)_R$
shifts $\psi$, and remains as a symmetry of the system.

In Refs.~\cite{Gauntetal,Bigetal} a one-parameter family of supergravity
solutions has
been found which describes five-branes wrapped on $\CC\PP^1$. The parameter,
denoted in Ref.~\cite{Gauntetal} by $k$, can take any real value. 
For $k<-1$ the solutions have
a naked singularity of a ``bad'' kind according to the criteria of
Ref.~\cite{MN2}, while for $k\geq 1$ the naked
singularity, although still present, is of a ``good'' kind. It is easy to
check that our solution coincides with the solution of 
Refs.~\cite{Gauntetal,Bigetal}
in the special case $k=-1$. The above-mentioned singularity is
at $Y=0,\rho=\frac{\pi}{2}.$ However, now it is clear that the singularity is 
an artefact of the supergravity approximation. 
The world-sheet CFT is manifestly non-singular. 

The authors of Ref.~\cite{Gauntetal} proposed that the solution with $k=-1$
represents the origin of the Coulomb branch of the five-brane theory, while
the solutions with $k>-1$ correspond to a one-dimensional submanifold of the
Coulomb branch. It is also proposed that the solutions with $k<-1$
 are unphysical. In Section~\ref{sec:CP1} we use a more general LSM to
derive the supergravity solution for general $k\geq -1$.
Our analysis
confirms the above interpretation of $k$, with some important clarifications.

Note that we have a well-defined world-sheet CFT only for $m>1$. This
is a signal that there is no decoupled theory on a single
five-brane~\cite{Witten,DS}.

Similarly, we can show that the six-dimensional linear dilaton background
corresponding to $n=3$ is related to NS five-branes wrapped on $\CC\PP^2$.
To this end we set $k=9m, m=2,3,\ldots,$ tensor with the $m^{\rm th}$
${\mathcal N}=2$ minimal model, and then take the $\Z_m$ orbifold.
The resulting $(2,2)$ superconformal field theory
has central charge $c=12.$ Therefore we can obtain a $(2,2)$ superstring
background by tensoring with a single free chiral superfield. To establish
a connection with wrapped five-branes, we perform T-duality along
the Killing vector field
$$
\frac{1}{m}\left(\frac{\partial}{\partial
\beta}+\frac{\partial}{\partial\psi}\right).
$$
The resulting eight-dimensional metric, B-field, and the dilaton have the
following form:
\begin{eqnarray}\label{SUGRAP2}
\dd s^2&=&\frac{g_3(Y)}{2} dY^2+2Y \dd s_{FS}^2 + m\dd\rho^2\\ \nn
&+& \left(1+\frac{9m}{2}\cot^2\rho\ g_3(Y)\right)^{-1}
\left(m\cot^2\rho\left(\dd\psi-3A\right)^2+
\frac{9}{2} m^2 g_3(Y)\dd\lambda^2\right),\\
B&=& \frac{m}{2\pi} \frac{\dd\lambda\wedge  
\left(\dd\psi-3A\right)}{1+\frac{9m}{2}\cot^2\rho\ g_3(Y)}, \\
\Phi&=&-\frac{Y}{3m}-\frac{1}{2}\log\left(\frac{9m}{2}\cos^2\rho+
\frac{\sin^2\rho}{g_3(Y)}\right)+const.
\end{eqnarray}
Here the variables $\psi$ and $\lambda$ have period $2\pi,$ and the 
function $g_3(Y)$ is given by
$$
g_3(Y)=\frac{2}{9m}\frac{\frac{1}{2}\left(\frac{2Y}{3m}\right)^2}{1-
\frac{2Y}{3m}+\frac{1}{2}\left(\frac{2Y}{3m}\right)^2-e^{-2Y/(3m)}}.
$$

At large $Y$, the background is an $\SS^3$ with H-flux
fibered over $\CP^2$. The fibration is associated with the
$U(2)$ frame bundle of $\CP^2$ and the $U(2)$ action
on $\SS^3$ determined by the twist we have chosen in
Section~\ref{sec:twist}.
Thus, we are dealing with a $\CP^2$ compactification of NS5-branes
with $(2,2)$ supersymmetry in $1+1$ dimensions.
The vector $U(1)$ R-symmetry is associated with the action (\ref{twistU1})
and shifts $\psi$. This remains a symmetry of the system.
On the other hand, the axial $U(1)$ R-symmetry,
identified as the action (\ref{RU1}), shifts $\lambda$ and is broken
to $\Z_{6m}$. Since in $(2,2)$ super-Yang-Mills the axial $U(1)$  
R-symmetry is anomaly free, this is somewhat puzzling.
A possible resolution will be discussed in Section~\ref{sec:decoup}.

This solution has a curvature
singularity at $Y=0,$ $\rho=\frac{\pi}{2}$.
There is
also a $\CC^3/\ZZ_3$ orbifold singularity at $Y=0,\rho=0$.
This is related
to the fact that the solution describes five-branes wrapped on the
exceptional divisor of the crepant resolution of a $\CC^3/\ZZ_3$
singularity. The K\"ahler class of the exceptional divisor is zero
for the above solution. A more general solution allowing a non-zero
size for the exceptional divisor is written down in Section~\ref{sec:CP2}.

One can also interpret the $n=4$ LSM with $k=16m$ tensored with
the $m^{\rm th}$ ${\mathcal N}=2$ minimal model as representing $m$ Euclidean 
NS five-branes wrapped on $\CC\PP^3$ in a Calabi-Yau 4-fold. 
We will not discuss this case any further in this paper.

\section{Mirror Description of Wrapped Five-Branes}\label{sec:mirror}

Let us recapitulate what we have seen.
We denote by
$\LSM_k^{(n)}$ the IR fixed point of
the linear sigma model (\ref{LSMnk}).
We first focus on the case $n=2$ which we denote simply
by $\LSM_k$.
We considered the ${\cal N}=2$ SCFT
\begin{equation}
{\mathcal C}_m
=\left[{\LSM_{4m}\over \Z_2}\times
{SU(2)_{m-2}\over U(1)}\right]/\Z^{\rm diag}_m,
\label{orb1}
\end{equation}
where ${SU(2)_{m-2}\over U(1)}$ stands for the level $(m-2)$
minimal model, $\Z_2$ corresponds to
the shift of ${\rm Im}\,P$ by half-period,
and $\Z^{\rm diag}_m$ is the diagonal subgroup of the
$U(1)\times \Z_m$ symmetry of the product theory.
The model $\LSM_{4m}$ has an asymptotic region which is described by
$\CC\PP^1\times \R\times U(1)_{4m}$ where $\R$ is the linear dilaton and
$U(1)_{4m}$ is the circle of radius $\sqrt{4m}$.
The $\Z_2$ orbifold group acts only on the $U(1)_{4m}$ factor,
resulting in $U(1)_{4m}/\Z_2=U(1)_m$.
Then using T-duality
$[U(1)_m\times {SU(2)_{m-2}\over U(1)}]/\Z_m\cong SU(2)_{m-2}$
\cite{OV,SKY},
we find that the asymptotic region describes 
$m$ NS5-branes wrapped on $\CC\PP^1$.
The theory (\ref{orb1}) can also be regarded as
a $\Z_{2m}$ orbifold of ${\rm LSM}\times {SU(2)\over U(1)}$,
and has the associated quantum symmetry $\Z_{2m}^{\rm qua}$.

To find the mirror description of ${\mathcal C}_m,$ we use the standard fact about 
orbifolds: the orbfold of a CFT ${\mathcal C}$ by some discrete symmetry
$G$ has a quantum symmetry $G'$ isomorphic to $G$,
 such that its orbifold reproduces the original CFT,
$({\mathcal C}/G)/G'\cong{\mathcal C}$.
Applying this to ${\mathcal C}_m$ in (\ref{orb1}) and
$G=(\Z_m\times\Z_m)/\Z_m^{\rm diag}\cong\Z_m$,
we find that ${\mathcal C}_m/G={\LSM_{4m}\over \Z_2}/\Z_m\times
{SU(2)_{m-2}\over U(1)}/\Z_m$, and therefore
\begin{equation}
{\mathcal C}_m\cong
\left[{\LSM_{4m}\over \Z_{2m}}\times
{SU(2)_{m-2}\over U(1)}/\Z_m\right]/\Z'_m,
\label{orb2}
\end{equation}
where we have used ${\LSM_{4m}\over \Z_2}/\Z_m\cong \LSM_{4m}/\Z_{2m}$.

It is known that the orbifold ${SU(2)_{m-2}\over U(1)}/\Z_m$
is mirror to the original minimal model, which can be realized as the
IR limit of the LG model of a single chiral superfield $X$
with the superpotential $W=X^m$.
The model $\LSM_{4m}/\Z_{2m}$ also
has a mirror description, as shown in Ref.~\cite{HK}.
It is given by two chiral superfields $Y$ and $Z$ of periodicity
$2\pi i$, with the following Kahler potential and superpotential
\begin{eqnarray}\label{K}
&&K={m\over 2}|Z|^2+a|Y|^2,\\ \label{W}
&&W=\e^{-mZ}(\e^{-Y}+\e^{Y}).
\end{eqnarray}
The parameter $a$ is mirror to the squashing of the metric of $\CC\PP^1$,
with the round metric corresponding to the degenerate limit $a\to 0$.
This mirror Landau-Ginzburg theory can be obtained as follows.
To incorporate the orbifold action, we must change the $P$-dependent part of
the action (\ref{LSMnk}) from ${k\over 4}(P+\overline{P}+V)^2$ to
${1\over 4m}(P+\overline{P}+2mV)^2$, where $P$
transforms as $P\to P+2mi\alpha$ under the gauge transformation. The period
of ${\rm Im}\,P$ is still $2\pi.$
Applying the method of Ref.~\cite{HV}, we find that the mirror is
a LG model of three chiral superfields $Y_1,Y_2,Y_P$,
all with period $2\pi i$, which are constrained by $Y_1+Y_2+2mY_P=0$.
The superpotential is given by $W=\e^{-Y_1}+\e^{-Y_2}$.
The K\"ahler potential is degenerate for a round $\CC\PP^1$ metric,
$K={m\over 2}|Y_P|^2$,
but can be made into a regular one
$K={m\over 2}|Y_P|^2+a|{Y_1-Y_2\over 2}|^2$ by squashing
$\CC\PP^1$.
We solve the constraint by setting $Y_1=Y+mZ$, $Y_2=-Y+mZ$, $Y_P=-Z$, 
and obtain the above K\"ahler potential and superpotential.

Thus, the mirror of the product theory 
${\LSM_{4m}\over \Z_{2m}}\times
{SU(2)_{m-2}\over U(1)}/\Z_m$ is a LG model
with the superpotential
\begin{equation}
W=\e^{-mZ}(\e^{-Y}+\e^{Y})+X^m.
\label{Wmir1}
\end{equation}
The theory ${\cal C}_m$ is therefore mirror to the orbifold of
this LG model with respect to the $\Z_m'$ discrete symmetry.
The orbifold group $\Z_m'$ acts on the fields as follows:
\begin{equation}
\Z_m':\,Y\to Y,\quad
Z\to Z-{2\pi i\over m},\quad
X\to \e^{{2\pi i\over m}}X.
\label{Zmir1}
\end{equation}
The action of the quantum symmetry $\Z_{2m}^{\rm qua}$ is
$Y\to Y+\pi i$, $Z\to Z+{\pi i\over m}$, $X\to X$.
We also note that the LG model has a dilaton represented by
the linear term in the supercurrent
${\cal J}|_{\rm linear}=(\partial_0-\partial_1){\rm Im}(Z)$.
Thus the dilaton is exactly linear:
\begin{equation}
\Phi=-{\rm Re}\,Z.
\label{dilaton}
\end{equation}
We see that the string coupling is weak in the asymptotic region
${\rm Re}\,Z \gg 0,$ while the strong coupling region
${\rm Re}\,Z \ll 0$ is effectively blocked by the Liouville-type
interaction.

Conformal field theories based on the
$SU(2)$ current algebra are classified by
simply laced Lie groups
\begin{eqnarray}
\Gamma_m&=&A_{m-1}\,\, \mbox{($m$ integer $\geq 2$)},\nn\\
&&D_{{m\over 2}+1}\,\, \mbox{($m$ even integer $\geq 6$)},\nn\\
&&E_6\,\,(m=12),\nn\\
&&E_7\,\,(m=18),\nn\\
&&E_8\,\, (m=30),
\end{eqnarray}
where $(m-2)$ is the level of the current algebra.
They have central charge $c=3(m-2)/m$ and
are denoted by $SU(2)_{\Gamma_m}$.
$SU(2)_{A_{m-1}}$ is the one we have been calling $SU(2)_{m-2}$
and corresponds to the system of $m$ ordinary NS5-branes.
Others corresponds to other types of NS5-branes.
For all $\Gamma_m$
we have the ${\mathcal N}=2$ supercoset
$SU(2)_{\Gamma_m}/U(1)$
which is called the ${\mathcal N}=2$ minimal model of type $\Gamma_m$.
These models arise as the IR fixed point of the LG models
with the following superpotentials \cite{Mart,VW}:
\begin{eqnarray}
&&W_{A_{m-1}}=X_1^m+X_2X_3,
\label{WA2}\\
&&W_{D_{{m\over 2}+1}}=X_1^{m\over 2}+X_1X_2^2+X_3^2,\\
&&W_{E_6}=X_1^4+X_2^3+X_3^2,\\
&&W_{E_7}=X_1^3X_2+X_2^3+X_3^2,\\
&&W_{E_8}=X_1^5+X_2^3+X_3^2.
\end{eqnarray}
They have discrete $\Z_m$ symmetry which acts on the LG fields by
\begin{equation}
X_i\to \e^{2\pi i q_i}X_i,
\end{equation}
where $q_i$ is the R-charge of $X_i$ such that
$W(\lambda^{q_i}X_i)=\lambda W(X_i)$.
Note that we have
\begin{equation}
q_1+q_2+q_3={m+1\over m}.
\end{equation}
As in the $A_{m-1}$ model, we have an identity
\begin{equation}
\left[U(1)_m\times {SU(2)_{\Gamma_m}\over U(1)}\right]/\Z_m
\cong SU(2)_{\Gamma_m}.
\end{equation}
Then it follows that $\Gamma_m$-type NS5-brane wrapped on $\CC\PP^1$ in
Eguchi-Hanson space is described by the CFT given by
(\ref{orb1}) or (\ref{orb2}), with $SU(2)_{m-2}$ replaced by
$SU(2)_{\Gamma_m}$.
Repeating the arguments leading to Eq.~(\ref{Wmir1}), we find that
the mirror is a LG orbifold with the superpotential
\begin{equation}
W=\e^{-mZ}(\e^{-Y}+\e^{Y})+W_{\Gamma_m},
\end{equation}
where the orbifold group $\Z_m'$ is generated by $Y\to Y$ and
\begin{equation}
Z\to Z-{2\pi i\over m},\quad
X_i\to \e^{2\pi i q_i}X_i.
\label{Zmact}
\end{equation}

Similarly, the theory describing $\Gamma_m$-type NS5-brane 
wrapped on $\CC\PP^2$ is given by
\begin{eqnarray}
{\mathcal C}^{(3)}_{\Gamma_m}
&=&\left[{\LSM_{9m}^{(3)}\over \Z_3}\times
{SU(2)_{\Gamma_m}\over U(1)}\right]/\Z^{\rm diag}_m
\label{orb3a}\\
&\cong&
\left[{\LSM_{9m}^{(3)}\over \Z_{3m}}\times
{SU(2)_{\Gamma_m}\over U(1)}/\Z_m\right]/\Z'_m.
\label{orb3}
\end{eqnarray}
The mirror is the LG model with the superpotential
\begin{equation}\label{LGCPtwo}
W=\e^{-mZ}(\e^{-Y_1}+\e^{-Y_2}+\e^{Y_1+Y_2})+W_{\Gamma_m},
\end{equation}
divided by the orbifold group $\Z_m'$ whose action on the fields is given by
$Y_i\to Y_i$ and Eq.~(\ref{Zmact}).
The theory has a quantum symmetry $\Z_{3m}^{\rm qua}$
and it acts on the mirror fields as
$Y_i\to Y_i+{2\pi i\over 3}$,
$Z\to Z-{2\pi i\over 3m}$, $X_i\to X_i$.

\section{Non-perturbative 4d Physics from the World-Sheet}\label{sec:CP1}

\subsection{Derivation of the Seiberg-Witten Solution}

The LG orbifold (\ref{Wmir1})-(\ref{Zmir1}) has a finite-dimensional
chiral ring spanned by $\Z_m'$-invariant operators
\begin{equation}
\e^{-(m-\ell)Z}X^{\ell},\quad
\ell = 0,1,\ldots,(m-2).
\end{equation}
All these operators are exactly marginal.
Let us deform the theory by adding these operators
to $W$:
\begin{equation}
W=\e^{-mZ}(\e^{-Y}+\e^Y)+X^m+\sum_{\ell=0}^{m-2}
u_{\ell}\,\e^{-(m-\ell)Z}X^{\ell}.
\label{deform}
\end{equation}
Since the parameters $u_{\ell}$ are exactly marginal,
Eq.~(\ref{deform}) represents a deformed string background.
Note that the deformation breaks the quantum symmetry
$\Z_{2m}^{\rm qua}$ to its subgroup.
As in Eq.~(\ref{WA2}),
we can add two extra variables $X_2$ and $X_3$
with the superpotential $X_2X_3$ without changing the IR fixed point.

Let us change the variables by setting
$X=\e^{-Z}\wtX$, $X_2=\e^{-mZ}\wtX_2$ and $X_3=\wtX_3$,
where the tilded variables are all invariant under
$\Z_m'$. 
Then $\e^{-mZ}$ factors out in the deformed superpotential:
\begin{equation}
W=\e^{-mZ}\left(
\e^{-Y}+\e^Y+\wtX^m+\sum_{\ell=0}^{m-2}u_{\ell}\,\wtX^{\ell}
+\wtX_2\wtX_3\right).
\end{equation}
We now show that this LG model is weakly equivalent
(in the sense of \cite{HV}) to the non-linear sigma model on
the non-compact Calabi-Yau 3-fold
\begin{equation}
\e^{-Y}+\e^Y+\wtX^m+\sum_{\ell=0}^{m-2}u_{\ell}\,\wtX^{\ell}
+\wtX_2\wtX_3=0.
\label{mirCY}
\end{equation}
Weak equivalence means that central charges of A-branes
in the two theories agree.
The central charge of an A-brane in the LG model
is measured by the period integral \cite{HV,HIV}
\begin{equation}
\Pi
=\int {1\over g_s}\dd Y\dd Z\dd X\dd X_2\dd X_3\exp\left(-W\right),
\label{Period}
\end{equation}
where $g_s$ is the string coupling, and the integration is over a
Lagrangian 5-cycle wrapped by the A-brane. This formula was derived
assuming that the dilaton is constant, but in the present theory
the dilaton is given by Eq.~(\ref{dilaton}), and thus the string coupling is
$g_s=\e^{\Phi_{\rm dilaton}}=\e^{-{\rm Re}\ Z}$. It is easy
to see how the expression (\ref{Period}) should be modified
so that it is holomorphic in all variables: one should simply replace
$g_s$ with $e^{-Z}.$
Then the period integral Eq.~(\ref{Period}) becomes
\begin{equation}
\Pi=\int
\e^Z\cdot \dd Y\dd Z\ \e^{-Z}\dd\wtX\e^{-mZ}\dd \wtX_2\dd\wtX_3
\exp\left(-W\right).
\end{equation}
We see that the Jacobian $\e^{-Z}$ from the change of
variables $X=\e^{-Z} \wtX$ is cancelled by the complexified string coupling.
This enables us to perform the $\e^{-mZ}$-integration which yields
\begin{equation}
\Pi=\int \dd Y\dd \wtX\dd \wtX_2\dd \wtX_3\,\delta\left(
\e^{-Y}+\e^Y+\wtX^m+\sum_{\ell=0}^{m-2}u_{\ell}\,\wtX^{\ell}
+\wtX_2\wtX_3\right).
\end{equation}
This is none other but the period integral
of the Calabi-Yau space (\ref{mirCY}), which measures the central 
charge of an A-brane.
It is well known~\cite{KLMVW,KKV} that the Type IIB superstring on
this space reduces in the low energy limit
to the Seiberg-Witten theory of the curve
\begin{equation}
\e^{-Y}+\e^Y+\wtX^m+\sum_{\ell=0}^{m-2}u_{\ell}\,\wtX^{\ell}=0,
\label{SWcurve}
\end{equation}
with the meromorphic differential 
$\lambda=\wtX\dd Y$. This is exactly the low energy effective theory
of the $SU(m)$ super-Yang-Mills, where $u_{\ell}$'s represent
the Coulomb branch vevs.
The breaking of the quantum symmetry $\Z_{2m}^{\rm qua}$
corresponds to the spontaneous breaking of the $\Z_{4m}$ R-symmetry.

This can be easily generalized to all ADE fivebranes.
The change of variables is $X_i=\e^{-mq_iZ}\wtX_i,$
and the holomorphic measure is given by
\begin{equation}
{1\over g_s}\dd Y\dd Z\prod_{i=1}^3\dd X_i
=\dd Y\e^{-mZ}\dd Z\prod_{i=1}^3\dd\wtX_i,
\end{equation}
where we used $\sum_{i=1}^3q_i=(m+1)/m$ and $g_s=\e^{-Z}$.
It follows that
the deformed system
is weakly equivalent to
the sigma model on the Calabi-Yau 3-fold
\begin{equation}
\e^{-Y}+\e^Y+W_{\Gamma_m}(\wtX_i,u_{\ell})=0,
\end{equation}
where $W_{\Gamma_m}(X_i,u_{\ell})$ is a versal deformation of
$W_{\Gamma_m}(X_i)$.
This is exactly the Calabi-Yau geometry that reproduces the
Seiberg-Witten effective theory of ${\mathcal N}=2$ super-Yang-Mills
with gauge group $\Gamma_m$~\cite{KLMVW}. The connection between the
five-brane theory and ${\mathcal N}=2$
super-Yang-Mills is discussed in more detail
in Section~\ref{sec:decoup}.

Given the equivalence of
the decoupled theory of Type IIB NS5-branes
and that of the ADE singularity in Type IIA,
the present system
is equivalent to Type IIA strings on
a non-compact Calabi-Yau manifold with an ADE singularity along $\CC\PP^1$.
In fact such systems are studied in the so called ``geometric
engineering'' approach \cite{KKV},
in which the Seiberg-Witten solution was also reproduced
using local mirror symmetry.
We also note that a transform from a similar LG model to the
CY geometry was performed in \cite{Lerche1}.

\subsection{Argyres-Douglas Points}

It is not easy to trace back what a generic deformation of $W$ does
to the wrapped five-brane system. However, there is a particular family
which has a simple but interesting meaning.
Consider the following one-parameter deformation that breaks
$\Z_{2m}^{\rm qua}$ to $\Z_m$:
\begin{equation}
W=\e^{-mZ}(\e^{-Y}+\e^Y)+X^m+\e^{t/2}\e^{-mZ}.
\end{equation}
That is, we set all $u_{\ell}$ to zero except $u_0=\e^{t/2}$.
By reversing the dualization process,
we find that this deformation corresponds to replacing ${\rm LSM}_{4m}/\Z_2$
in (\ref{orb1}) by the IR fixed point of the following linear sigma model.
The gauge group is $U(1)\times U(1),$ and there are
four chiral superfields with the charges described
in the table~(\ref{table}).
\begin{equation}
\begin{array}{ccccc}
    &\Phi_1&\Phi_2& Q & P\\
U(1)&  1   &  1   &0  &{\rm shift}\\ 
U(1)&  1   &  1   &-2& 0
\end{array}
\label{table}
\end{equation}
We have added one field $Q$ and one $U(1)$ gauge group
to the basic linear sigma model for ${\rm LSM}_{4m}$.
The orbifold group $\Z_{m}^{\rm diag}$ acts
on the minimal model as before and on the LSM as
$\Phi_i\to\Phi_i,Q\to \e^{2\pi i\over m}Q$ and $P\to P$.
The breaking $\Z_{2m}^{\rm qua}\to\Z_m$ corresponds to the
fact that the final system can only be regarded as
a $\Z_{m}$ orbifold.

The new $U(1)$ has a non-trivial Fayet-Iliopoulos-theta parameter
which is equal to $t$.
One can easily see that the complex geometry of the Higgs branch is that
of a resolved $A_1$ singularity, or Eguchi-Hanson space, where $t$ is 
roughly (up to world-sheet instanton corrections) the size of $\CC\PP^1$. This
is discussed in greater detail below.
The original undeformed CFT --- the Landau-Ginzburg without the
$\e^{t/2}\e^{-mZ}$-term --- corresponds to $t\to-\infty$, the orbifold
limit of the Eguchi-Hanson space.
The other limit $t\to +\infty$ is the large volume limit. Formally, this
can be seen as follows.
Making a change of variables $Z=Z'+t/(2m)$ and taking the strict $t\to+\infty$
limit, we find the superpotential
\begin{equation}
W=\e^{-mZ'}+X^m.
\end{equation}
This is indeed the mirror dual of $m$ flat NS5-branes~\cite{GKP,GK}.

For generic $t$ the infrared limit of this LSM is a non-singular CFT,
which can be thought of as a non-linear sigma-model of the Higgs branch
obtained by integrating out the vector multiplets.
But for special values of $t$ it may happen that
the effective twisted superpotential for
the field-strength superfield $\Sigma=\overline{D}_+D_-V$
vanishes; for these values of $t$,
a ``Coulomb branch'' develops
and the CFT becomes singular~\cite{Wittenphases}.
In our model the effective twisted superpotential is
\footnote{We can neglect the first $U(1)$ since the corresponding vector
multiplet is always heavy via mixing with $P$.}
\begin{equation}
W_{\it eff}
=-t\Sigma-\sum_{i=1}^3Q_i\Sigma(\log(Q_i\Sigma)-1)
=(-t+2\log(-2))\Sigma,
\end{equation}
where $Q_i$ stands for the charge vector $(1,1,-2)$.
We see that the system is singular exactly at one point
\begin{equation}
t=\log 4.
\end{equation}

Let us see what happens to the mirror theory at this point.
Since $\e^{t/2}=\sqrt{\e^{\log 4}}=\pm 2$,
the SW curve (\ref{SWcurve}) at this point is
$\e^{-Y}+\e^{Y}+\wtX^m\pm 2=0$.
At $\e^{-Y}=\mp 1$ and $\wtX=0$,
the curve has an $A_{m-1}$ singularity
\begin{equation}
Y^2+\wtX^m=0.
\end{equation}
This is nothing but the curve at the
Argyres-Douglas point in the $SU(m)$ super-Yang-Mills
\cite{AD,APSW,EHIY} if $m>2$, and it is a massless monopole point
if $m=2$. The world-sheet CFT becomes singular
precisely when some particles become massless in the
space-time theory
 --- from the string theory point of view, these particles
are wrapped D-branes.
This is the phenomenon which we already encountered before. For example,
a massless charged hypermultiplet
at Type IIB conifold singularity \cite{Stro}
and enhanced gauge symmetry (massless W-boson)
at Type IIA $ADE$ singularity without B-field
\cite{Aspin}.
Note that our LSM description is in Type IIA string theory
and the singularity is the $A_1$ case of \cite{Aspin}.
However, by tensoring with the minimal model and taking an orbifold,
we see the phenomenon similar to
\cite{Stro} (it is equivalent for $m=2$).

This analysis of course generalizes to NS5-branes of type $\Gamma_m$,
in which case we get Argyres-Douglas points of ${\mathcal N}=2$
super-Yang-Mills theory with gauge group $\Gamma_m$. 
In Ref.~\cite{EH} it was suggested that 
certain
${\mathcal N}=2$ $d=4$ SCFTs fall into the ADE classification. 
The present discussion relates
this to the ADE classification of
NS5-branes, which in turn is equivalent to
the ADE classification of $SU(2)$ modular invariants \cite{CIZ}.

\subsection{Supergravity Description of the Coulomb Branch}

It is not difficult to find the supergravity solution corresponding
to the above one-parameter family of CFTs. We start with the LSM realization
and integrate out the vector multiplets classically. This yields
a non-linear sigma-model with a K\"ahlerian target space. The metric is
\begin{equation}\label{defclasmet}
\dd s^2=2f_s(r)dr^2
+\frac{r^2}{2f_s(r)}\left(\dd\psi-\cos\theta \dd\phi\right)^2+
\frac{r^2}{2}\left(\dd\theta^2+\sin^2\theta \dd\phi^2\right),
\end{equation}
where
$$
f_s(r)=\frac{3r^2-2s}{2(r^2-s)}+\frac{2r^2}{k},\quad s={\rm Re}\ t.
$$
The variable $\psi$ has period $2\pi.$ The variable $r$ ranges from $\sqrt s$ to $+\infty$ if $s>0$, and from $0$ to $+\infty$ if $s\leq 0.$

The first thing to note about this metric is that it is non-singular
for $s>0$ and has a curvature singularity at $r=0$ for $s\leq 0.$
This means that for $s\leq 0$ quantum effects on the world-sheet are large,
and one cannot use the non-linear sigma-model to describe the infrared
physics of the LSM. In fact, even if $s$ is positive but small, the
curvature is large near $r=\sqrt s.$ Thus the semi-classical treatment of
the world-sheet is valid only for $s$ large and positive. We therefore
restrict ourselves to this regime.

Second, it is easy to see that topologically, and even algebro-geometrically, 
the manifold with the above metric is a copy of $T^*\CC\PP^1$.
The parameter $s$ is proportional to the area of the holomorphic 2-sphere
which generates the second homology of $T^*\CC\PP^1$.

Third, the above metric is not Ricci flat, so to determine the IR
limit of the LSM we must figure out what it flows to. Since the
target space is $T^*\CC\PP^1,$ the first thing that comes to mind
is that the end-point of the flow is either a Eguchi-Hanson space,
or a two-center Taub-NUT space, the two well-known Ricci flat 
metrics on $T^*\CC\PP^1.$ However, neither of them can be the
end-point of the RG flow for our LSM. The Taub-NUT metric has
a $U(1)\times U(1)$ isometry, while our metric (and therefore the
whole RG trajectory) has a $U(2)$ isometry. The Eguchi-Hanson metric
has the right isometry, and in fact near the zero section the above metric
is identical to Eguchi-Hanson. But their behaviors at infinity are quite
different: while the Eguchi-Hanson metric is asymptotically locally
Euclidean (ALE), our metric is not. Rather, for fixed $r$ we have
a squashed $\RR\PP^3$, with the squashing increasing with $r$. In this
respect our metric resembles the two-center Taub-NUT metric, except that
the curvature of our metric falls off slower than in the Taub-NUT case. 

Since the Eguchi-Hanson metric is the only K\"ahlerian Ricci flat 
metric with $U(2)$ isometry, we conclude that the end-point cannot
be Ricci flat. Rather, it is described by a K\"ahlerian dilatonic
solution to the beta-function equations. Using the method of Ref.~\cite{KKL},
it is easy to show that the most general such solution with $U(2)$ isometry and the asymptotics as in Eq.~(\ref{defclasmet}) is given by
Eqs.~(\ref{ntwoIRmetric}) and Eq.~(\ref{ntwoIRdilaton}), but with 
$g_2(Y)$ replaced with
\begin{equation}\label{sugraclmb}
\tilde{g}_2(Y)=\frac{1}{2m}\frac{\frac{Y}{m}}{e^{-(Y-\sigma)/m}
\left(1-\frac{\sigma}{m}\right)-1+\frac{Y}{m}}.
\end{equation}
In addition, $Y$ now ranges from $\sigma$ to $+\infty.$

The parameter $\sigma>0$ measures the area of $\CC\PP^1$. 
Since the periods of the
K\"ahler form are not affected by the RG flow in world-sheet perturbation theory, we must set $\sigma=s.$ 
As discussed above, this derivation is valid for
$s\gg 0.$ Of course, world-sheet instantons can correct the relation
between $s$ and $\sigma$, but these corrections are exponentially small for
large $s.$

After tensoring with an ${\mathcal N}=2$ minimal model and performing T-duality,
one gets a six-dimensional SUGRA background which is identical to the
one found in Refs.~\cite{Gauntetal,Bigetal}. The parameter $k$ in
Ref.~\cite{Gauntetal} is related to our $s$ as follows:
$$
k=\left(\frac{s}{m}-1\right)\exp\left(\frac{s}{m}\right).
$$ 
This confirms that changing $k$ corresponds to moving on the Coulomb
branch of the five-brane theory. However, supergravity is
a good description everywhere only for $k$ large and positive, i.e. when
the size of $\CC\PP^1$ is large. Here ``good description'' has the following meaning: although the supergravity solution has a singularity for all
$k$, one can perform a T-duality which converts it to an (orbifold of)
the product of an ${\mathcal N}=2$ minimal model and a smooth four-dimensional
dilatonic background. It is the latter background which can be treated in the
supergravity approximation.

As discussed above, there is also a good ``global'' interpretation of
the solution with $k=-1$: it corresponds to five-branes at the origin of the Coulomb branch. For other values of $k$ supergravity can be used only away
from the singular locus containing five-branes. 

\section{Compactification on $\CC\PP^2$}\label{sec:CP2}

In this section we discuss Type IIB NS five-branes wrapped
on $\CC\PP^2$ in a similar manner. 
As discussed above, the naive expectation is that the low-energy theory
reduces to $(2,2)$ super-Yang-Mills. 

The quantity we would like to compute is again the mass of A-branes in the LG 
model Eq.~(\ref{LGCPtwo}). It is given by the following period integral:
$$
\Pi
=\int e^Z \dd Y_1 \dd Y_2 \dd Z\dd X\dd X_2\dd X_3\exp\left(-W\right).
$$
Making a change of variables
$$
X_i=\e^{-mq_iZ}\wtX_i,\quad i=1,2,3,
$$
and integrating over $Z,$ we find that the period integral is the same
as for a Calabi-Yau 4-fold defined by the equation
\begin{equation}
\e^{-Y_1}+\e^{-Y_2}+\e^{Y_1+Y_2}+W_{\Gamma_m}(\wtX_i)=0.
\label{CY4}
\end{equation}
This was also obtained in Ref.~\cite{Lerche,Kaste}
as the local mirror of a Calabi-Yau 4-fold with ADE singularity along
$\CC\PP^2$. 

While in the 4d case the mass of BPS states was interpreted in terms of
a prepotential, in the 2d case one can interpret it in terms of an
effective superpotential~\cite{Lerche,Kaste,GVW}. The BPS states
we are discussing are D4-branes wrapped on special Lagrangian 4-cycles
in a Calabi-Yau 4-fold $X$.
In the low-energy 2d field theory they should be thought of as kinks 
interpolating between different vacua. As explained in Ref.~\cite{GVW},
the vacua are labelled by the periods of the 4-form flux $G$ on $X$. Indeed,
since a D4-brane is magnetically charged with respect to $G$,
the VEV of $G$ jumps as one moves across the D4-brane. The jump is
equal to the Poincar\'{e} dual of the cycle wrapped by the brane.

Possible choices of $G$ are constrained by the requirement of $(2,2)$ supersymmetry~\cite{BB}. In the low-energy field theory these
constraints can be interpreted as arising from F-terms
which depend on complex and K\"ahler moduli~\cite{GVW}. 
Since complex structure moduli and K\"ahler moduli live in chiral and
twisted chiral multiplets, respectively, there will be a superpotential
depending on $G$ and complex structure moduli and a twisted superpotential
depending on $G$ and K\"ahler moduli. We are only concerned with the former. 
It is given by
$$
W=\frac{1}{2\pi}\int \Omega\wedge G,
$$
where $\Omega$ is the holomorphic 4-form on the Calabi-Yau (it is defined
up to a constant factor). In our case $\Omega$ is given by
$$
\wtX_1 \dd Y_1 \dd Y_2 \dd \wtX_2 \dd \wtX_3.
$$
Then it is easy to see that the BPS central charge of a special 
Lagrangian brane is given by the change in this superpotential as one goes across the corresponding kink. 

If we undo mirror symmetry, K\"aler and complex structure moduli, as
well as ordinary and twisted superpotentials, 
are exchanged. Thus the superpotential we have written down
maps to a twisted superpotential for the low-energy theory of
Type IIB NS five-branes wrapped on $\CC\PP^2.$ This
twisted superpotential is induced by world-sheet
instanton effects. One may ask if this twisted superpotential can
be interpreted in terms of $(2,2)$ super-Yang-Mills, or is it
an LST effect. This issue is discussed in the next section. 

Finally, let us consider
the following one-parameter family of deformations
of the CY geometry (\ref{CY4}) 
$$
\e^{-Y_1}+\e^{-Y_2}+\e^{Y_1+Y_2}+W_{\Gamma_m}(\widetilde{X}_i)=\mu.
$$
It corresponds to replacing
${\rm LSM}_{9m}^{(3)}/\Z_3$ in (\ref{orb3a})
by the IR limit of the following gauge theory
\begin{equation}
\begin{array}{cccccc}
    &\Phi_1&\Phi_2&\Phi_3& Q & P\\
U(1)&  1   &  1   &  1   & 0 &{\rm shift}\\ 
U(1)&  1   &  1   &  1   &-3 & 0
\end{array}
\nn
\end{equation}
with the orbifold group $\Z_m^{\rm diag}$ acting as $\Phi_i\to\Phi_i,
Q\to\e^{2\pi i\over m}Q$ and $P\to P$.
The deformation parameter $\mu$ corresponds to $\e^{t/3}$
where $t$ is the FI-Theta parameter of the second $U(1)$. 
One can write down the corresponding
one-parameter family of supergravity solutions generalizing 
Eq.~(\ref{SUGRAP2}).
It is obtained by replacing $g_3(Y)$ with 
$$
{\tilde g}_3(Y)=\frac{Y^2}{3} 
\frac{e^{2Y/(3m)}}{\int_\sigma^Y t^2 e^{2t/(3m)} dt},
$$
where $\sigma>0,$ and restricting the range of $Y$ to $[\sigma,+\infty).$
This family describes a one-dimensional submanifold of the Coulomb branch
defined by the equations $u_1=\ldots=u_\ell=0.$ One can also think
of it as describing $m$ NS5-branes wrapping the exceptional divisor
of a noncompact Calabi-Yau 3-fold which is a crepant resolution of
a $\CC^3/\ZZ_3$ singularity. Note that taking 
$\sigma>0$ resolves the $\ZZ_3$ orbifold singularity present in 
Eq.~(\ref{SUGRAP2}).

\section{Relation to Super-Yang-Mills Theories}\label{sec:decoup}

The supergravity solution~(\ref{SUGRAP1}) preserves $1/4$ 
supersymmetry~\cite{Gauntetal}, so the effective field theory in four 
dimensions has ${\mathcal N}=2$ supersymmetry. What is this theory?
The naive approach is to replace the theory of NS5-branes with its
low-energy limit, ${\mathcal N}=2$ $d=6$ super-Yang-Mills theory, 
and then perform the Kaluza-Klein reduction. This procedure yields
${\mathcal N}=2$ $d=4$ super-Yang-Mills with the same gauge group
as in six dimensions.
However, since the decoupled theory on the NS5-branes is a non-local
Little String Theory rather than a field theory, care is needed to
make sense of this result. A difficulty arises when one tries to compute
the 4d gauge coupling. Naively, it is given by
$$
g_4^2=\frac{g_6^2}{4\pi R^2},
$$
where $4\pi R^2$ is the area of $\CC\PP^1$. But the area of $\CC\PP^1$ in
the supergravity solution depends on $Y$, so one must specify at which
$Y$ it must be evaluated.
Usually, we measure the size at infinity in space, $Y=+\infty$,
but then the area of $\CC\PP^1$ is infinite too,
which gives zero 4d coupling.
Another way to put the question is to ask
what is the size of $\CC\PP^1$ on which the LST is compactified. In ordinary
field theory, there is always a definite answer, but in the present case,
since $R(Y)\ra +\infty$ for $Y\ra \infty,$ the situation is less clear.

Following Ref.~\cite{MN}, we
interpret the $Y$-dependence of various quantities as
the dependence on the energy scale $E$ in the following way:
\begin{equation}\label{UVIR}
E\sim \e^{-\Phi(Y)}.
\end{equation}
This relation is motivated by the way the mass of the W-boson scales with
$g_{st}=e^{\Phi}$.
We would like to have
a region of $Y$ that corresponds to high energies in
4d physics but low energy compared to string scale and Kaluza-Klein scale,
where the 4d gauge coupling is small.
Since the 6d gauge coupling is
$g_6^2=2(2\pi)^3$ and the volume of $\CC\PP^1$ is
$4\pi R^2=2\pi Y$, the 4d gauge coupling at $Y$ is
$g_4^2=2(2\pi)^3/2\pi Y=2(2\pi)^2/Y$.
It is small at large $Y$ where the dilaton behaves as
$\Phi(Y)\simeq -Y/2m+\Phi_0$.
In such a region, the 4d gauge coupling depends on the
energy scale $E$ as
\begin{equation}
{8\pi^2\over g_4^2}\sim Y\sim
2m\log(E/\e^{-\Phi_0}).
\end{equation}
This is the correct running of the gauge coupling, with the 4d dimensional
transmutation scale being
\begin{equation}
\Lambda=\e^{-\Phi_0}.
\end{equation}
To extract just the super-Yang-Mills, we need $\Lambda$
to be smaller than the string scale (which is $1$)
and the Kaluza-Klein scale
(which is of order $1/4\pi R^2\sim 1/2\pi Y$).
This requires $e^{\Phi_0}\gg 1$.
Thus the string coupling is extremely large
in the relevant region of small $Y$.

Alternatively, we may go away from the origin of
the Coulomb branch, so that the logarithmic running of $g_4^2$ stops
at the scale of the Coulomb branch VEV $v$.
This means that we must consider
the one-parameter family of supergravity solutions~(\ref{sugraclmb}).
We identify the Coulomb branch VEV with 
the mass of a W-boson which is represented by a
D-brane.
Obviously, this D-brane must be sitting at the locus where its mass
is minimized and the string coupling is maximized, i.e. at $Y=\sigma.$ 
Then its mass is going to be of order $\exp(-\Phi(\sigma)).$
The argument above can be repeated in this set-up:
energy $E$ is replaced by the VEV $v\sim \exp(-\Phi(\sigma)),$ 
and we end up with the same conclusion that decoupling requires
$e^{\Phi_0}\gg 1$.

Thus the theory of wrapped five-branes reduces to pure ${\mathcal N}=2$
super-Yang-Mills only when the string coupling is taken to be very large.
In this regime our world-sheet description is not applicable, and one has
to use the S-dual description in terms of D5-branes wrapped on $\CC\PP^1$.
This is completely analogous to Ref.~\cite{MN}. On the other hand, if
one insists on working at weak string coupling, then the gauge theory
scale $\Lambda$ is comparable or larger than the string scale, and one 
is dealing with an LST in 4d, rather than with a gauge theory. 
Our world-sheet computation of the prepotential in Section~\ref{sec:CP1} 
nevertheless agrees on the nose with the Seiberg-Witten solution of 
${\mathcal N}=2$ super-Yang-Mills. This happens because of a supersymmetric 
non-renormalization theorem, which states that the $N=2$ $d=4$ prepotential
cannot depend on the VEVs of hypermultiplets. In our case, the dilaton is
part of a hypermuliplet, and therefore the prepotential cannot depend on it.

One can try to perform a similar analysis for five-branes wrapped on
$\CC\PP^2,$
although it is complicated by infrared divergences which are present
already at the classical level. The classical mass of the W-boson on the
Coulomb branch is made of a finite contribution of order $v$ and a
a divergent self-energy of order $g_2^2 L,$ where $L$ is the infrared cut-off.
The mass of a D-brane has a similar structure, with the constant
contribution of order $\exp(-\Phi(\sigma)).$ We see that to keep
the regularized D-brane mass below the string scale, we must work
in the regime $e^{\Phi(\sigma)}\gg 1.$ 

Finally, let us comment on the size of world-sheet instantons in the
decoupling limit. In the $\CC\PP^1$ case, world-sheet instantons
precisely correspond to gauge theory instantons.
This is because in the S-dual (D5-brane) picture
string world-sheet becomes the world-volume of a D1-brane, and D1-branes bound to D5-branes are nothing but Yang-Mills instantons.
Their effects are bound to survive the decoupling limit.
This was explicitly demonstrated in Ref.~\cite{KKV} in the geometric 
engineering picture.
In the $\CC\PP^2$ case, it appears that world-sheet instanton
effects disappear in the decoupling limit. For simplicity, let us set
$m=2$. Then according to Ref.~\cite{Lerche} one-instanton contribution
is of order
$$
v^{-6} e^{-\frac{c}{g_2}}.
$$
The mass scale $c$ is of order one in string units; its precise
value is unimportant.
We want to take the 
decoupling limit $g_2\ra 0, v\ra 0,$ while keeping the strength of the
2d interactions fixed. This means we must keep $g_2/v$ fixed. Clearly, in
this limit world-sheet instanton effects disappear. Another way to argue
this is to note~\cite{Lerche} that the decoupled theory is unchanged
if we replace $\CC\PP^2$ by any other Hirzebruch surface. On the other
hand, such a replacement will drastically change world-sheet instanton
contributions.
The disappearance of the world-sheet instanton effects in the decoupling limit
seems to solve the puzzle mentioned in Section~\ref{sec:T},
regarding the breaking of the
axial R-symmetry $U(1)_A$ dow to $Z_{6m}$.

In the 4-fold description of wrapped branes, it is easy to see that there
are other effects which could lift the Coulomb branch.
These are five-brane instantons wrapping both the $\CC\PP^2$ base and
the $A_{m-1}$ fiber. Similar five-brane instantons in M-theory compactified 
on a Calabi-Yau 4-fold are known to be responsible for the generation
of the Affleck-Harvey-Witten superpotential in ${\mathcal N}=2$ $d=3$ super-Yang-Mills~\cite{WittenW,KV}. Since IIA on a 4-fold can
be obtained from M-theory on a 4-fold by compactifying an extra
circle, it is clear that five-brane instantons in Type IIA
at finite string coupling also induce a superpotential on the
Coulomb branch. It is not clear if such effects survive in the
strict decoupling limit. An analogous question in field theory is
whether the Affleck-Harvey-Witten superpotential survives
dimensional reduction to 2d.

It is interesting to note that while world-sheet instantons
induce a twisted superpotential depending on the K\"ahler moduli,
five-brane instantons induce an ordinary superpotential for
the T-duals of the K\"ahler moduli. Therefore it is impossible to 
write down a local action which includes both kinds of instantons. 
This is not so surprising, since five-brane instantons and world-sheet instantons are related by electric-magnetic duality and therefore are 
mutually non-local.

\section{Concluding Remarks}\label{sec:conclusion}

In this paper, we provided a world-sheet description of
NS5-branes wrapped on $\CP^1$ in Eguchi-Hanson space and
on $\CP^2$ in a non-compact CY 3-fold.

A striking aspect of the construction
is its simplicity.
Consider the origin of the moduli space of
the $\CP^1$ compactification.
Except for the well-understood minimal model,
the non-trivial part of the world-sheet theory is
the IR limit of a very simple linear sigma model
--- $U(1)$ gauge theory with two charge 1 matter fields plus a
field $P$ that shifts under gauge transformations.
Without the field $P$, the linear sigma model corrsponds to
the non-linear sigma model
on $\CP^1$ which is asymptotically free
(the FI-theta parameter is dimensionally transmuted to a holomorphic
scale parameter $\Lambda$)
and has a mass gap.
Introduction of the field $P$ corresponds to
promoting the scale parameter $\Lambda$ to a field, since
the dual of $P$ plays the role of the dynamical FI-Theta parameter.
In other words, ${\rm Re}\,P$ plays the role of the dynamical Weyl mode
or ``Liouville mode'', which would appear when we quantize
2d gravity coupled to this massive sigma model.
The field ${\rm Im}\,P$ corresponds to adding a circle fibered over
$\CP^1$.
One may say that the present system is the world-sheet theory
of a non-critical superstring on
$\RR^{3+1}\times [\CP^1
\tilde{\times} \SS^1\times {SU(2)_{m-2}\over U(1)}]/\ZZ_m$,
where $\tilde{\times}$ stands for a non-trivial fibration.
With the orbifold action taken into account,
the product $\SS^1\times  {SU(2)_{m-2}\over U(1)}$
can be replaced by $SU(2)_{m-2}$. Thus we are considering the
non-critical superstring on
\begin{equation}
\RR^{3+1}\times \CP^1\,\tilde{\times}\, SU(2)_{m-2}.
\end{equation}
The fibration $\CP^1\tilde{\times} SU(2)_{m-2}$
is the one associated with the ${\mathcal N}=2$ twisting.
We have shown that this describes the
$\CP^1$ compactification of $m$ NS5-branes,
at the origin of the moduli space. 
Moving away from the origin of the moduli space
corresponds to adding massive
matter, or performing a massive deformation of the minimal model
${SU(2)_{m-2}\over U(1)}$.
In the mirror LG model description it is more apparent that
the scale parameter is promoted to a field.

The K\"ahlerian dilatonic backgrounds discovered in Ref.~\cite{KKL}
and used in this paper as building blocks, are very natural generalizations
of the 2d Euclidean black hole and share many properties with the latter.
For example, we have seen that they all arise as IR fixed points of an
infinite sequence of linear sigma models. Since the 2d black hole is
integrable, it is natural to ask if its higher-dimensional generalizations
also posess this property. We do not have strong evidence either against
or in favor of this conjecture. We only performed the following integrability
test. We have derived above the mirror description of our CFTs
in terms of Landau-Ginzburg models. Given a Landau-Ginzburg model, it is 
quite straightforward to check whether it has non-trivial B\"acklund
transformations depending on fields and their derivatives up to a certain
order, see e.g. Ref.~\cite{ZIS}. 
Therefore we tried to look for B\"acklund
transformations in mirror LG models, with fermionic fields set to zero. 
In the case of the 2d black hole the mirror LG model is a super-Liouville theory, whose integrability and B\"acklund transformations are well known.
The LG model mirror to the 4d generalization of the 2d black hole is
given by Eqs.~(\ref{K}-\ref{W}) with $a=0.$ Unfortunately, since the K\"ahler
potential is degenerate, the only way to make sense of this model
is to regard it as a limit of the same model with $a\neq 0.$ We have checked
that in the neighborhood of $a=0$ the LG model defined by
Eqs.~(\ref{K}-\ref{W})
does not have non-trivial B\"acklund transformations, if we allow dependence on
derivatives up to fourth order. 

\section*{Acknowledgments}
We would like to thank Tohru Eguchi,
Jaume Gomis, Aki Hashimoto and David Tong for useful discussions,
and Juan Maldacena 
for explanations about the decoupling limit for wrapped five-branes.
K.~H. was supported in part by NSF-DMS 0074329
and by NSF-PHY 0070928. 
A.~K. was supported in part by DOE grant DE-FG03-92-ER40701.


\newpage

\begin{thebibliography}{99}

\bibitem{tHooft}
G.~'t Hooft,
``A Planar Diagram Theory For Strong Interactions'',
Nucl.\ Phys.\ B {\bf 72} (1974) 461.

\bibitem{Polyakov}
A.~M.~Polyakov,
``String theory and quark confinement'',
Nucl.\ Phys.\ Proc.\ Suppl.\  {\bf 68} (1998) 1
[arXiv:hep-th/9711002];
``The wall of the cave'',
Int.\ J.\ Mod.\ Phys.\ A {\bf 14} (1999) 645
[arXiv:hep-th/9809057].

\bibitem{LST}
N.~Seiberg,
``New theories in six dimensions
and matrix description of M-theory on  T**5 and T**5/Z(2),''
Phys.\ Lett.\ B {\bf 408} (1997) 98
[arXiv:hep-th/9705221].


\bibitem{CHS}
C.~G.~Callan, J.~A.~Harvey and A.~Strominger,
``World Sheet Approach To Heterotic Instantons And Solitons,''
Nucl.\ Phys.\ B {\bf 359} (1991) 611;
``Supersymmetric string solitons'',
arXiv:hep-th/9112030.


\bibitem{ABKS}
O.~Aharony, M.~Berkooz, D.~Kutasov and N.~Seiberg,
``Linear dilatons, NS5-branes and holography,''
JHEP {\bf 9810}, 004 (1998) [arXiv:hep-th/9808149].

\bibitem{OV}
H.~Ooguri and C.~Vafa,
``Two-Dimensional Black Hole and Singularities of CY Manifolds'',
Nucl.\ Phys.\ B {\bf 463} (1996) 55
[arXiv:hep-th/9511164].



\bibitem{BH}
E.~Witten, ``String theory and black holes,'' 
Phys.\ Rev.\ D {\bf 44}, 314 (1991);

G.~Mandal, A.~M.~Sengupta, and A.~R.~Wadia, 
``Classical solutions of two-dimensional string theory,''
Mod.\ Phys.\ Lett.\ A {\bf 6}, 1685 (1991);

S.~Elitsur, A.~Forge, and E.~Rabinovici, ``Some global aspects of
string compactifications,'' Nucl.\ Phys.\ B {\bf 359}, 581 (1991).


\bibitem{GKP}
A.~Giveon, D.~Kutasov and O.~Pelc,
``Holography for non-critical superstrings,''
JHEP {\bf 9910}, 035 (1999) [arXiv:hep-th/9907178];


\bibitem{GK}
A.~Giveon and D.~Kutasov,
``Little string theory in a double scaling limit,''
JHEP {\bf 9910}, 034 (1999)
[arXiv:hep-th/9909110];
``Comments on double scaled little string theory,''
JHEP {\bf 0001}, 023 (2000)
[arXiv:hep-th/9911039].

\bibitem{ES}
T.~Eguchi and Y.~Sugawara,
``Modular invariance in superstring on Calabi-Yau n-fold with
A-D-E  singularity,''
Nucl.\ Phys.\ B {\bf 577} (2000) 3
[arXiv:hep-th/0002100].


\bibitem{MN} 
J.~M.~Maldacena and C.~Nunez,
``Towards the large N limit of pure N = 1 super Yang Mills,''
Phys.\ Rev.\ Lett.\  {\bf 86}, 588 (2001)
[arXiv:hep-th/0008001].

\bibitem{Gauntetal}
J.~P.~Gauntlett, N.~Kim, D.~Martelli and D.~Waldram,
``Wrapped fivebranes and N = 2 super Yang-Mills theory,''
Phys.\ Rev.\ D {\bf 64}, 106008 (2001) [arXiv:hep-th/0106117]

\bibitem{Bigetal}
F.~Bigazzi, A.~L.~Cotrone and A.~Zaffaroni,
``N = 2 gauge theories from wrapped five-branes,''
Phys.\ Lett.\ B {\bf 519}, 269 (2001) [arXiv:hep-th/0106160].

\bibitem{GR}
J.~Gomis and J.~G.~Russo,
``D = 2+1 N = 2 Yang-Mills theory from wrapped branes,''
JHEP {\bf 0110}, 028 (2001)
[arXiv:hep-th/0109177].

\bibitem{Gauntetal2}
J.~P.~Gauntlett, N.~w.~Kim, D.~Martelli and D.~Waldram,
``Fivebranes wrapped on SLAG three-cycles and related geometry,''
JHEP {\bf 0111}, 018 (2001)
[arXiv:hep-th/0110034].

\bibitem{KKL}
E.~Kiritsis, C.~Kounnas and D.~Lust,
``A Large Class Of New Gravitational And Axionic Backgrounds For Four-Dimensional Superstrings,''
Int.\ J.\ Mod.\ Phys.\ A {\bf 9}, 1361 (1994)
[arXiv:hep-th/9308124].


\bibitem{HK}
K.~Hori and A.~Kapustin,
``Duality of the fermionic 2d black hole and N = 2 Liouville theory
as mirror symmetry,'' JHEP {\bf 0108}, 045 (2001)
[arXiv:hep-th/0104202].

\bibitem{HV}
K.~Hori and C.~Vafa,
``Mirror symmetry'', arXiv:hep-th/0002222.

\bibitem{CIZ}
A.~Cappelli, C.~Itzykson and J.~B.~Zuber,
``Modular Invariant Partition Functions In Two-Dimensions'',
Nucl.\ Phys.\ B {\bf 280} (1987) 445;
``The ADE Classification Of Minimal And A1(1) Conformal Invariant Theories,''
Commun.\ Math.\ Phys.\  {\bf 113} (1987) 1;

D.~Gepner and Z.~Qiu,
``Modular Invariant Partition Functions For Parafermionic Field Theories'',
Nucl.\ Phys.\ B {\bf 285} (1987) 423;

A.~Kato,
``Classification Of Modular Invariant Partition Functions
In Two-Dimensions'',
Mod.\ Phys.\ Lett.\ A {\bf 2} (1987) 585.

\bibitem{Calabi}
E.~Calabi, ``M\'{e}triques k\"ahl\'{e}riennes et fibr\'{e}s
holomorphes,'' Ann.\ Sci.\ \'{E}cole Norm.\ Sup.\ {\bf 12} (1979) 269.

\bibitem{SKY}
F.~Ravanini and S.~K.~Yang,
``Modular Invariance In N=2 Superconformal Field Theories'',
Phys.\ Lett.\ B {\bf 195} (1987) 202;

Z.~Qiu,
``Modular Invariant Partition Functions For
N=2 Superconformal Field Theories,''
Phys.\ Lett.\ B {\bf 198} (1987) 497.


\bibitem{MMS}
J.~Maldacena, G.~W.~Moore and N.~Seiberg,
``Geometrical interpretation of D-branes in gauged WZW models,''
JHEP {\bf 0107}, 046 (2001)
[arXiv:hep-th/0105038].

\bibitem{MN2}
J.~Maldacena and C.~Nunez,
``Supergravity description of field theories on
curved manifolds and a no go theorem,'' 
Int.\ J.\ Mod.\ Phys.\ A {\bf 16}, 822 (2001)
[arXiv:hep-th/0007018].


\bibitem{Witten}
E.~Witten,
``On the conformal field theory of the Higgs branch,''
JHEP {\bf 9707}, 003 (1997)
[arXiv:hep-th/9707093].

\bibitem{DS}
D.~E.~Diaconescu and N.~Seiberg,
``The Coulomb branch of (4,4) supersymmetric field theories in two  dimensions,''
JHEP {\bf 9707}, 001 (1997)
[arXiv:hep-th/9707158].



\bibitem{Mart}
E.~J.~Martinec,
``Algebraic Geometry And Effective Lagrangians'',
Phys.\ Lett.\ B {\bf 217} (1989) 431.


\bibitem{VW}
C.~Vafa and N.~P.~Warner,
``Catastrophes And The Classification Of Conformal Theories'',
Phys.\ Lett.\ B {\bf 218} (1989) 51.


\bibitem{HIV}
K.~Hori, A.~Iqbal and C.~Vafa,
``D-branes and mirror symmetry'', arXiv:hep-th/0005247.


\bibitem{KLMVW}
A.~Klemm, W.~Lerche, P.~Mayr, C.~Vafa and N.~P.~Warner,
``Self-Dual Strings and N=2 Supersymmetric Field Theory'',
Nucl.\ Phys.\ B {\bf 477} (1996) 746
[arXiv:hep-th/9604034].



\bibitem{KKV}
S.~Katz, A.~Klemm and C.~Vafa,
``Geometric engineering of quantum field theories,''
Nucl.\ Phys.\ B {\bf 497}, 173 (1997) [arXiv:hep-th/9609239].

\bibitem{Lerche1}
W.~Lerche, ``On a boundary CFT description of nonperturbative N = 2 
Yang-Mills  theory,'' arXiv:hep-th/0006100.


\bibitem{Wittenphases}
E.~Witten,
``Phases of N = 2 theories in two dimensions,''
Nucl.\ Phys.\ B {\bf 403}, 159 (1993)
[arXiv:hep-th/9301042].

\bibitem{AD}
P.~C.~Argyres and M.~R.~Douglas,
``New phenomena in SU(3) supersymmetric gauge theory'',
Nucl.\ Phys.\ B {\bf 448} (1995) 93
[arXiv:hep-th/9505062].

\bibitem{APSW}
P.~C.~Argyres, M.~Ronen Plesser, N.~Seiberg and E.~Witten,
``New N=2 Superconformal Field Theories in Four Dimensions'',
Nucl.\ Phys.\ B {\bf 461} (1996) 71
[arXiv:hep-th/9511154].

\bibitem{EHIY}
T.~Eguchi, K.~Hori, K.~Ito and S.~K.~Yang,
``Study of ${\mathcal N}=2$ Superconformal Field Theories in $4$ Dimensions'',
Nucl.\ Phys.\ B {\bf 471} (1996) 430
[arXiv:hep-th/9603002].



\bibitem{Stro}
A.~Strominger, ``Massless Black Holes and Conifolds in String Theory,''
Nucl.\ Phys.\ B {\bf 451} (1995) 96 
[arXiv:hep-th/9504090].

\bibitem{Aspin}
P.~S.~Aspinwall,
``Enhanced gauge symmetries and K3 surfaces'',
Phys.\ Lett.\ B {\bf 357} (1995) 329
[arXiv:hep-th/9507012].

\bibitem{EH}
T.~Eguchi and K.~Hori,
``N = 2 superconformal field theories in 4 dimensions and A-D-E  classification,'' arXiv:hep-th/9607125.



\bibitem{Lerche}
W.~Lerche,
``Fayet-Iliopoulos potentials from four-folds,''
JHEP {\bf 9711}, 004 (1997)
[arXiv:hep-th/9709146].


\bibitem{Kaste}
P.~Kaste,
``On the twisted chiral potential in 2d and the analogue of rigid special  geometry for 4-folds,''
JHEP {\bf 9906}, 021 (1999)
[arXiv:hep-th/9904218].

\bibitem{GVW}
S.~Gukov, C.~Vafa and E.~Witten,
``CFT's from Calabi-Yau four-folds,''
Nucl.\ Phys.\ B {\bf 584}, 69 (2000)
[Erratum-ibid.\ B {\bf 608}, 477 (2000)]
[arXiv:hep-th/9906070].







\bibitem{BB}
K.~Becker and M.~Becker,
``M-Theory on Eight-Manifolds,''
Nucl.\ Phys.\ B {\bf 477}, 155 (1996)
[arXiv:hep-th/9605053].










\bibitem{WittenW}
E.~Witten,
``Non-Perturbative Superpotentials In String Theory,''
Nucl.\ Phys.\ B {\bf 474}, 343 (1996)
[arXiv:hep-th/9604030].

\bibitem{KV}
S.~Katz and C.~Vafa,
``Geometric engineering of N = 1 quantum field theories,''
Nucl.\ Phys.\ B {\bf 497}, 196 (1997)
[arXiv:hep-th/9611090].

\bibitem{ZIS}
A.~V.~Ziber, N.~H.~Ibragimov, and A.~B.~Shabat, ``Equations of
Liouville type,'' Soviet\ Math.\ Dokl.\ {\bf 20}, 1183 (1979).

\end{thebibliography}
\end{document}